\documentclass{aa}
\usepackage{comment}
\usepackage{amssymb}

\usepackage{changepage}
\usepackage{graphicx}
\usepackage{subfig}
\usepackage{hyperref}
\usepackage{txfonts}
\usepackage{natbib}
\DeclareGraphicsExtensions{.pdf,.png,.jpg,.ps,.tif}
\bibpunct{(}{)}{;}{a}{}{,} 
\begin{document} 

  \title{Star-forming brightest cluster galaxies at $z\sim0.4$ in KiDS.\\ Further studies of cold gas and stellar properties.}
  \titlerunning{Star-forming brightest cluster galaxies at $z\sim0.4$ in KiDS. Further studies of cold gas and stellar properties.}

   \author{G. Castignani
          \inst{1,2}\fnmsep\thanks{e-mail: gianluca.castignani@unibo.it}
          \and
          M. Radovich\inst{3}
        \and
          F. Combes\inst{4,5}
          \and
          P. Salom\'e\inst{4}
          \and
          L. Moscardini\inst{1,2,6}
          \and
          S. Bardelli\inst{2}
          \and
          C. Giocoli\inst{2,6}          
          \and
          G. Lesci\inst{1,2}
          \and
          F. Marulli\inst{1,2,6}
          \and
          M. Maturi\inst{7,8}
          \and
          E. Puddu\inst{9}
          \and 
          M. Sereno\inst{2,6}
          \and
          D. Tramonte\inst{10,11,12}
          }
   \institute{Dipartimento di Fisica e Astronomia ''Augusto Righi'', Alma Mater Studiorum Università di Bologna, Via Gobetti 93/2, I-40129 Bologna, Italy   
   \and
   INAF - Osservatorio  di  Astrofisica  e  Scienza  dello  Spazio  di  Bologna,  via  Gobetti  93/3,  I-40129,  Bologna,  Italy
   \and
   INAF - Osservatorio Astronomico di Padova, vicolo dell'Osservatorio 5, I-35122 Padova, Italy
   \and
   Observatoire de Paris, LERMA, CNRS, Sorbonne University, PSL Research Universty, 75014 Paris, France 
   \and
   Coll\`{e}ge de France, 11 Place Marcelin Berthelot, 75231 Paris, France
   \and INFN - Sezione di Bologna, viale Berti-Pichat 6/2, I-40127 Bologna, Italy
   \and  Center for Astronomy - University of Heidelberg, Albert-Ueberle-Stra{\ss}e 2, 69120 Heidelberg, Germany 
   \and Institute of Theoretical Physics - University of Heidelberg, Albert-Ueberle-Stra{\ss}e 2, 69120 Heidelberg, Germany 
   \and
   INAF - Osservatorio di Capodimonte, Salita Moiariello 16, 80131 Napoli, Italy      
   \and
   Department of Physics, Xi'an Jiaotong-Liverpool University, 111 Ren'ai Road,  Suzhou Dushu Lake Science and Education Innovation District, Suzhou Industrial Park, Suzhou 215123, P.R. China
   \and
   Purple Mountain Observatory, No. 8 Yuanhua Road, Qixia District, Nanjing 210034, China
   \and
    NAOC-UKZN Computational Astrophysics Center (NUCAC), University of Kwazulu-Natal, Durban, 4000, South Africa
     }
\date{Received: November 4, 2022; Accepted: February 10, 2023}

  \abstract
   {Brightest cluster galaxies (BCGs) at the centers of clusters are among the most massive galaxies in the Universe.    Their star formation history and stellar mass assembly are highly debated. Recent studies suggest the presence of an emerging population of intermediate-$z$ star-forming and gas-rich BCGs, whose molecular gas reservoirs that feed star formation might be impacted by strong environmental processing.    We have selected three of the most strongly star-forming $z\sim0.4$ BCGs in the equatorial field of the Kilo-Degree Survey (KiDS) and observed them with the IRAM 30m telescope in the first three CO transitions.    We found clear double-horn CO(1$\rightarrow$0) and CO(3$\rightarrow$2) emission for the KiDS~1433 BCG, yielding a large molecular gas reservoir with $M_{H_2}=(5.9\pm1.2)\times10^{10}~M_\odot$ and a high gas-to-stellar mass ratio $M_{H_2}/M_\star=(0.32^{+0.12}_{-0.10})$. We thus increase the still limited sample of distant BCGs with detections in multiple CO transitions. The double-horn emission for the KiDS~1433 BCG implies a low gas concentration, while a modeling of the spectra yields an extended molecular gas reservoir, with a characteristic radius of $\sim$(5-7)~kpc, which is reminiscent of the mature extended-disk phase that is observed in some local BCGs.
   For the remaining two BCGs, we are able to set robust upper limits of $M_{H_2}/M_\star<0.07$ and $<0.23$, which are among the lowest for distant BCGs. We then combined our observations with available stellar, star formation, and dust properties of the targeted BCGs, and compared them with a sample of $\sim100$ distant cluster galaxies, including additional intermediate-$z$ BCGs, with observations in CO from the literature.    
   Altogether, our analysis shows that the molecular gas properties of star-forming BCGs are heterogeneous. On the one hand, gas-rich BCGs show extended gas reservoirs that sustain the significant star formation activity, but the efficiency is low, which is reminiscent of recent gas infall.  On the other hand, the existence of similarly star forming but gas-poor BCGs suggests that gas depletion precedes star formation quenching.}
   
   \maketitle

\begin{figure*}[th!]\centering
\captionsetup[subfigure]{labelformat=empty}
\subfloat[]{\hspace{0.cm}\includegraphics[trim={0.5cm 0cm 3cm 
0cm},clip,width=0.32\textwidth,clip=true]{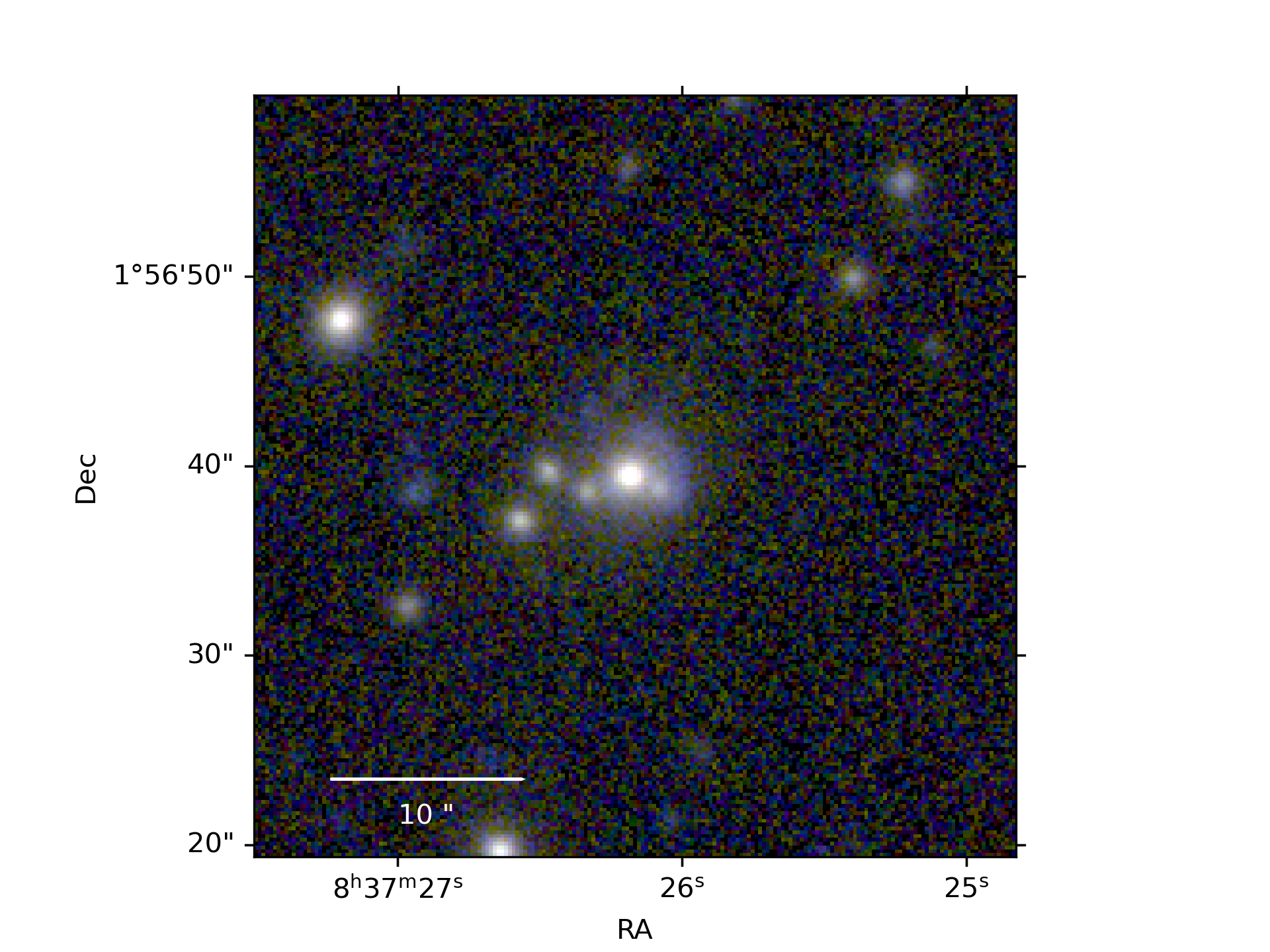}}
\hspace{0.1cm}\subfloat[]{\hspace{0.cm}\includegraphics[trim={0.5cm 0cm 3cm 
0cm},clip,width=0.32\textwidth,clip=true]{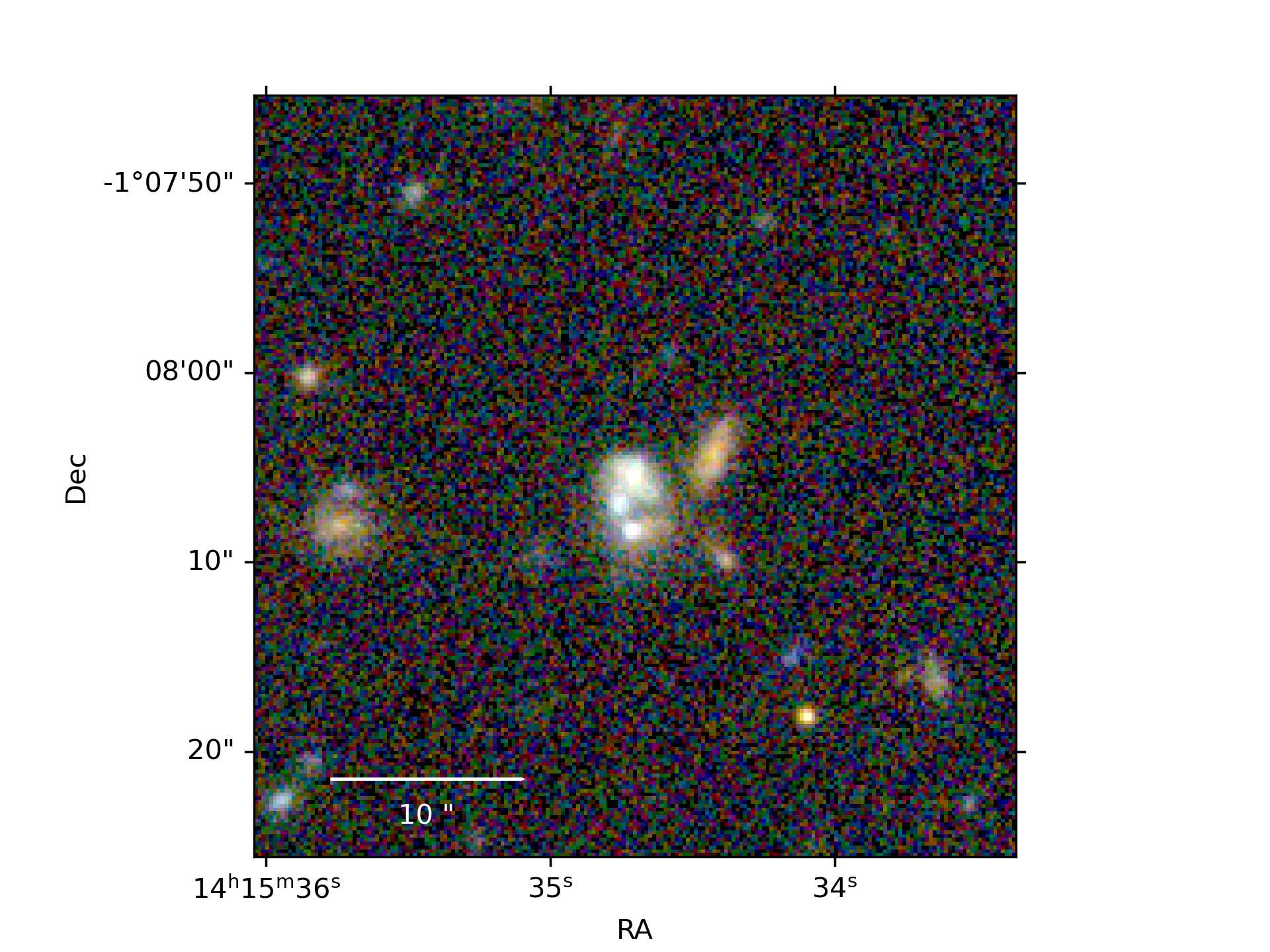}}
\hspace{0.1cm}\subfloat[]{\hspace{0.cm}\includegraphics[trim={0.5cm 0cm 3cm 
0cm},clip,width=0.32\textwidth,clip=true]{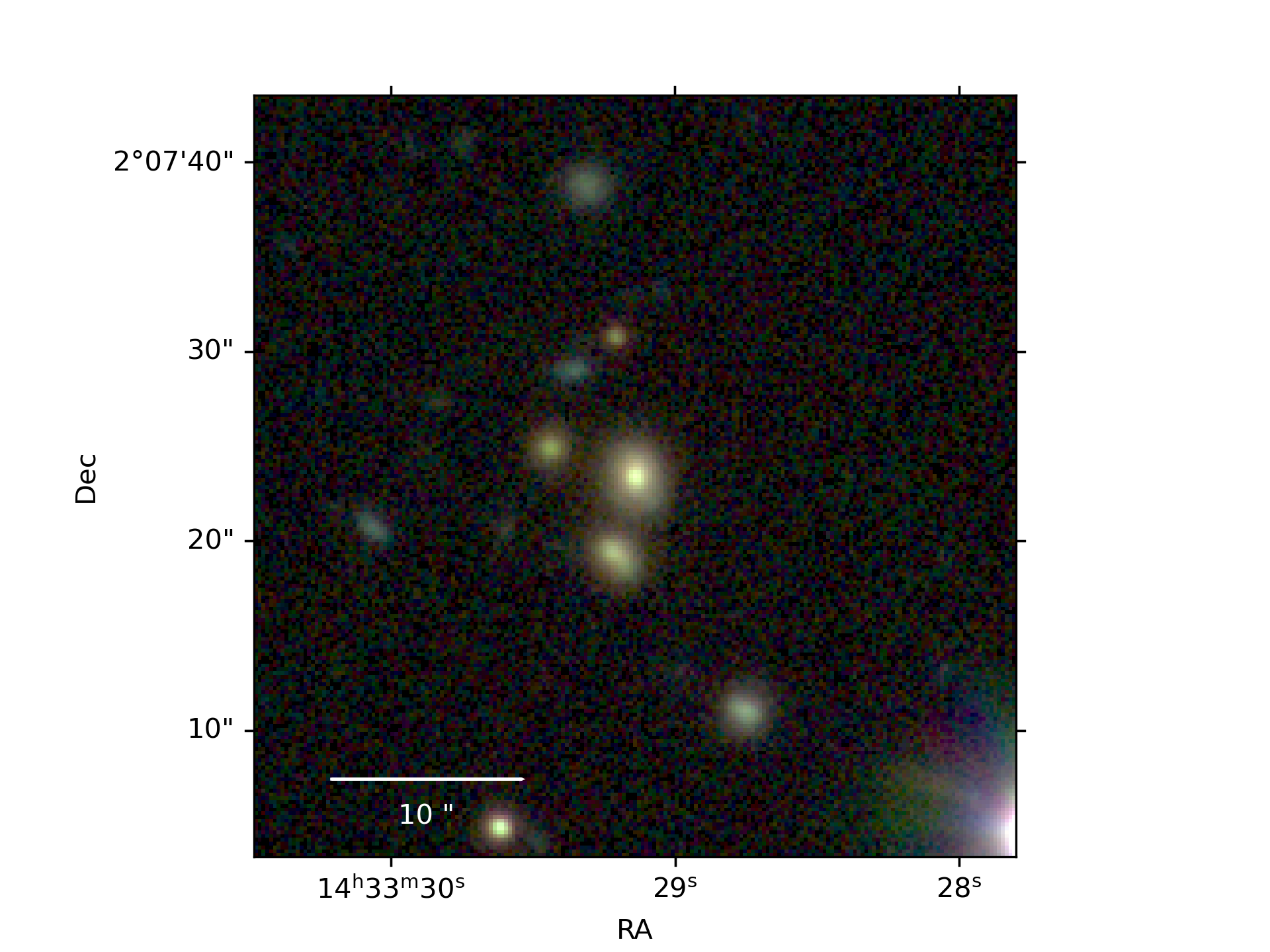}}
\caption{Color-composite $40^{\prime\prime}\times40^{\prime\prime}$ KiDS images (red~=~\textsf{i} band, green~=~\textsf{r} band, and blue~=~\textsf{g} band) with OmegaCAM at the VST, 
centered around the BCGs KiDS~0837 (left), KiDS~1415 (center), and KiDS~1433 (right). North is up,  and east is to the left.}\label{fig:BCG_images}
\end{figure*}
%

\section{Introduction}\label{sec:introduction}
Brightest cluster galaxies (BCGs) are optimal laboratories for studying the effect of the cluster environment on galaxy evolution. At $z\sim0,$ they are almost invariably associated with passively evolving (cD) ellipticals at the centers of clusters \citep{Lauer2014}. To explain their exceptional masses and luminosities, BCGs are
thought to evolve via environment-driven mechanisms such as dynamical friction \citep{White1976}, galactic cannibalism \citep{Hausman_Ostriker1978}, interactions with the intracluster medium \citep{Stott2012}, and 
cooling flows \citep{Salome_Combes2003}, regulated by active galactic nucleus (AGN) feedback \citep[see, e.g.,][for some reviews] {Fabian2012,Miley_DeBreuck2008,Magliocchetti2022}.


Simulations predict that a large fraction of the BCG stellar mass is assembled at high redshifts ($z
\sim3-5$), while smaller galaxies are later swallowed by the BCG itself: a process that allows the BCGs to grow in mass and size \citep{DeLucia_Blaizot2007}. BCGs indeed double their stellar mass since
$z\sim1$ \citep{Lidman2012}, which is consistent with evolution via dry accretion of satellites \citep{Collins2009,Stott2011}. However, recent studies have proposed that high star formation and large molecular gas reservoirs are present in
a number of intermediate-redshift BCGs \citep{Fogarty2015,Fogarty2019,Castignani2020a, Castignani2022a,Dunne2021,OSullivan2021}, where star formation is fed by rapid gas deposition, possibly induced by recent mergers or interactions with the companions \citep{Castignani2022a}. 
These results  imply strong environmental quenching mechanisms for distant BCGs. Therefore, the standard scenario of an early formation of BCGs, with a peak of star formation activity followed by passive evolution, needs to be reconsidered, at least for the subpopulation of these distant BCGs with significant ongoing star formation activity (i.e., with a star formation rate, SFR, $\gtrsim20~M_\odot$/yr) and with large molecular ($H_2$) gas reservoirs ($M_{H_2}\gtrsim10^{10}~M_\odot$).

In recent studies, we have started a large campaign exploiting millimeter facilities with the final goal {of understanding} the impact of the cluster environments in processing the galactic gas reservoirs, with a particular focus on distant ($z>0.3$) BCGs, their mass assembly, and  their star formation properties \citep{Castignani2019,Castignani2020a,Castignani2020b,Castignani2020c,Castignani2022a,Castignani2022b}. The advent of ongoing and forthcoming infrared-to-optical surveys such as the Dark Energy Survey \citep{McClintock2019,Aguena2021}, 
the Vera Rubin Observatory Legacy Survey of Space and Time \citep[LSST;][]{Ivezic2019}, and {\it Euclid} \citep{Scaramella2022} will revolutionize the field in the next years, with thousands of clusters up to $z\sim2$, bringing an unprecedented leverage for constraining the evolution of BCGs in clusters \citep{Rhodes2017}.

These surveys also include the multiwavelength Kilo Degree Survey \citep[KiDS;][]{deJong2017,Kuijken2019}. In its final release, this imaging survey will cover about 1,350~deg$^2$ of the sky \citep{deJong2017,Kuijken2019}, equally divided between an equatorial (KiDS-N) and a southern field (KiDS-S). KiDS exploits VLT Survey Telescope (VST) \textsf{ugri} optical observations  and Visible and Infrared Survey Telescope for Astronomy (VISTA) telescope \textsf{ZYJHK} infrared imaging, and it is able to detect sources down to a limiting AB magnitude  $\textsf{r}\sim25.0$  ($5\sigma$ in $2^{\prime\prime}$ aperture). Optical-infrared photometry enables the determination of accurate photometric redshifts up to $z\sim1$ \citep{Benitez2000,Hildebrandt2021}. These are complemented by spectroscopy, as KiDS is also covered by {the} Galaxy And Mass Assembly \citep[GAMA;][]{Baldry2010} and the Sloan Digital Sky Survey \citep[SDSS;][]{York2000} spectroscopic surveys. This rich data set has enabled the detection of a sample of $\sim$8000 galaxy clusters up to $z\sim0.8$ in KiDS at a signal-to-noise ratio (S/N) $>3$ \citep{Maturi2019,Bellagamba2019} with the Adaptive Matched Identifier of Clustered Objects (AMICO) cluster finder \citep{Bellagamba2018}, which searches for overdensities of galaxies using photometric redshifts with a matched filtering.

In this work, we study three star-forming BCGs that we selected as those among the most star-forming BCGs in KiDS.  We present our analysis resulting from new observations of the three BCGs in the first three CO transitions with the Institut de Radioastronomie Millimétrique (IRAM) 30m telescope. {We follow a similar approach to that of our recent study, \citet{Castignani2022a}, hereafter denoted C22a,} where we investigated the molecular gas and star formation properties of three additional BCGs drawn from the same parent sample. 
In C22a, we were able to double the number of distant BCGs with clear detections in at least two CO lines based on new molecular gas observations. To explain the global stellar, gas, and star formation properties of the BCGs, we suggested that a substantial amount of the molecular gas might have been accreted by the BCGs, possibly via accretion by the nearby cluster core companions, but might still not have been efficiently converted into stars.
{Alternatively, the observed molecular gas reservoirs could  originate from the cooling of the hot X-ray atmospheres around the BCGs.} We also showed, for the first time for distant BCGs, that gas excitation is {correlated} with the specific star formation rate (sSFR) and the star formation efficiency at which molecular gas is consumed, as was previously found only in other types of star-forming galaxies. Highly excited gas is present only in cool-core and highly star-forming (SFR$\gtrsim100~M_\odot$/yr) BCGs, where the low-entropy cool cores of clusters sustain high levels of star formation and favor the condensation and excitation of cold gas in the BCGs \citep{Castignani2020a}. Together, these results suggest that environmentally driven mechanisms play an important role in processing the molecular gas that feeds star formation in distant BCGs. In order to better understand the physical mechanisms that cause this processing, it is thus essential to increase the still limited sample of distant BCGs with detections at different molecular gas transitions, which probe different gas densities, and possibly different degrees of environmental processing.

The paper is structured as follows. In Sect.~\ref{sec:BCGsample} we {introduce} the BCGs that are the subject of this work, in Sect.~\ref{sec:observations_and_data_reduction} we describe the molecular gas observations and analysis, in Sect.~\ref{sec:results} we present the results, and in Sect.~\ref{sec:conclusions} we summarize them and draw our conclusions. Stellar mass and SFR estimates reported in this work for the three targeted BCGs rely on the \citet{Chabrier2003} initial mass function (IMF). Magnitudes are reported in the AB system. Throughout this work, we adopt a flat {$\Lambda$-cold dark matter} ($\Lambda$CDM) cosmology with matter density $\Omega_{\rm m} = 0.30$, dark energy density $\Omega_{\Lambda} = 0.70$, and Hubble constant $h=H_0/(100\, \rm km\,s^{-1}\,Mpc^{-1}) = 0.70$. 

\begin{figure}[]\centering
\captionsetup[subfigure]{labelformat=empty}
\subfloat[KiDS~0837]{\hspace{0.cm}\includegraphics[trim={0cm 0cm 0cm 
0cm},clip,width=0.5\textwidth,clip=true]{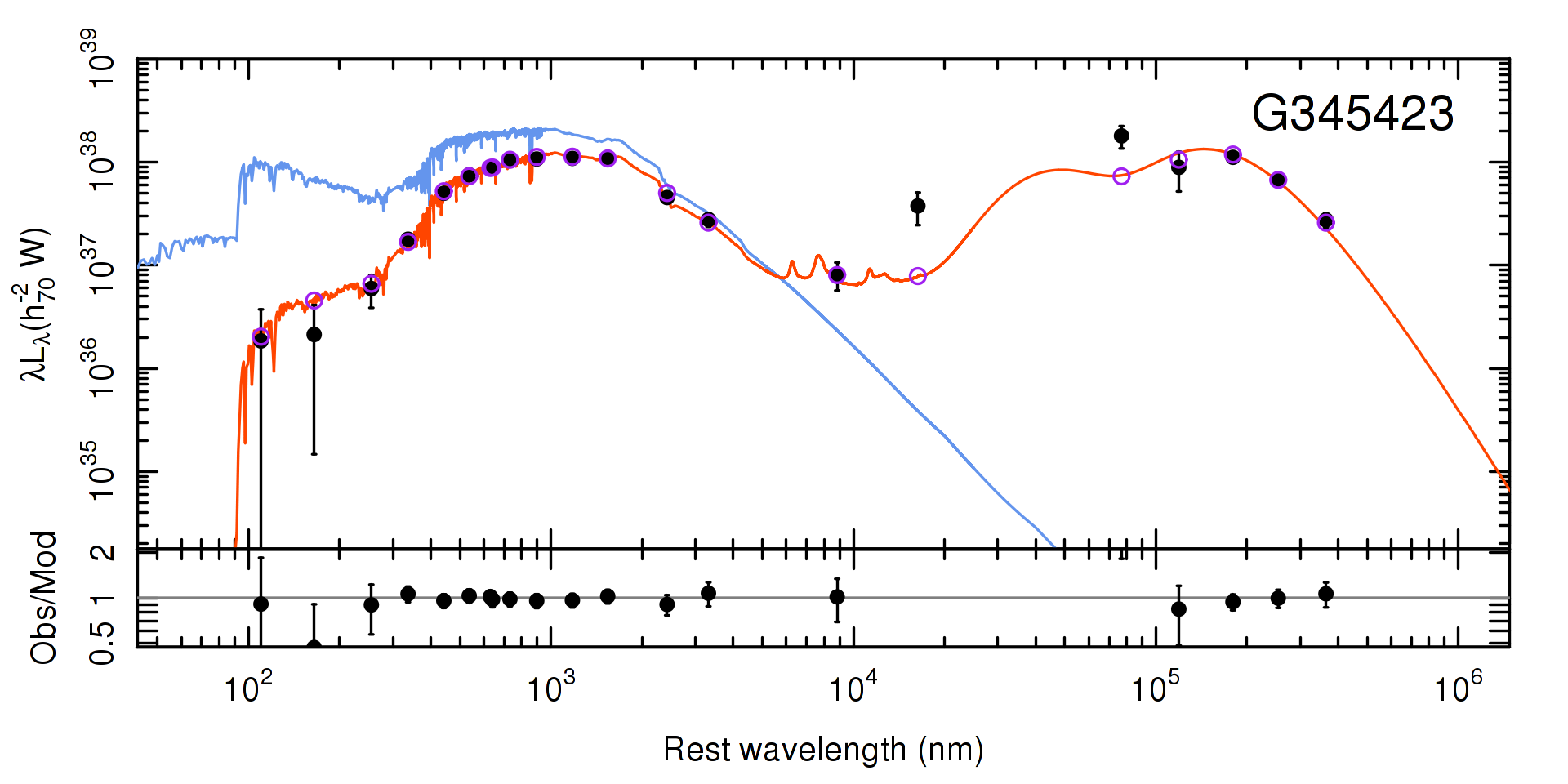}}
\hspace{0.2cm}\subfloat[KiDS~1415]{\hspace{0.cm}\includegraphics[trim={0cm 0cm 0cm 
0cm},clip,width=0.5\textwidth,clip=true]{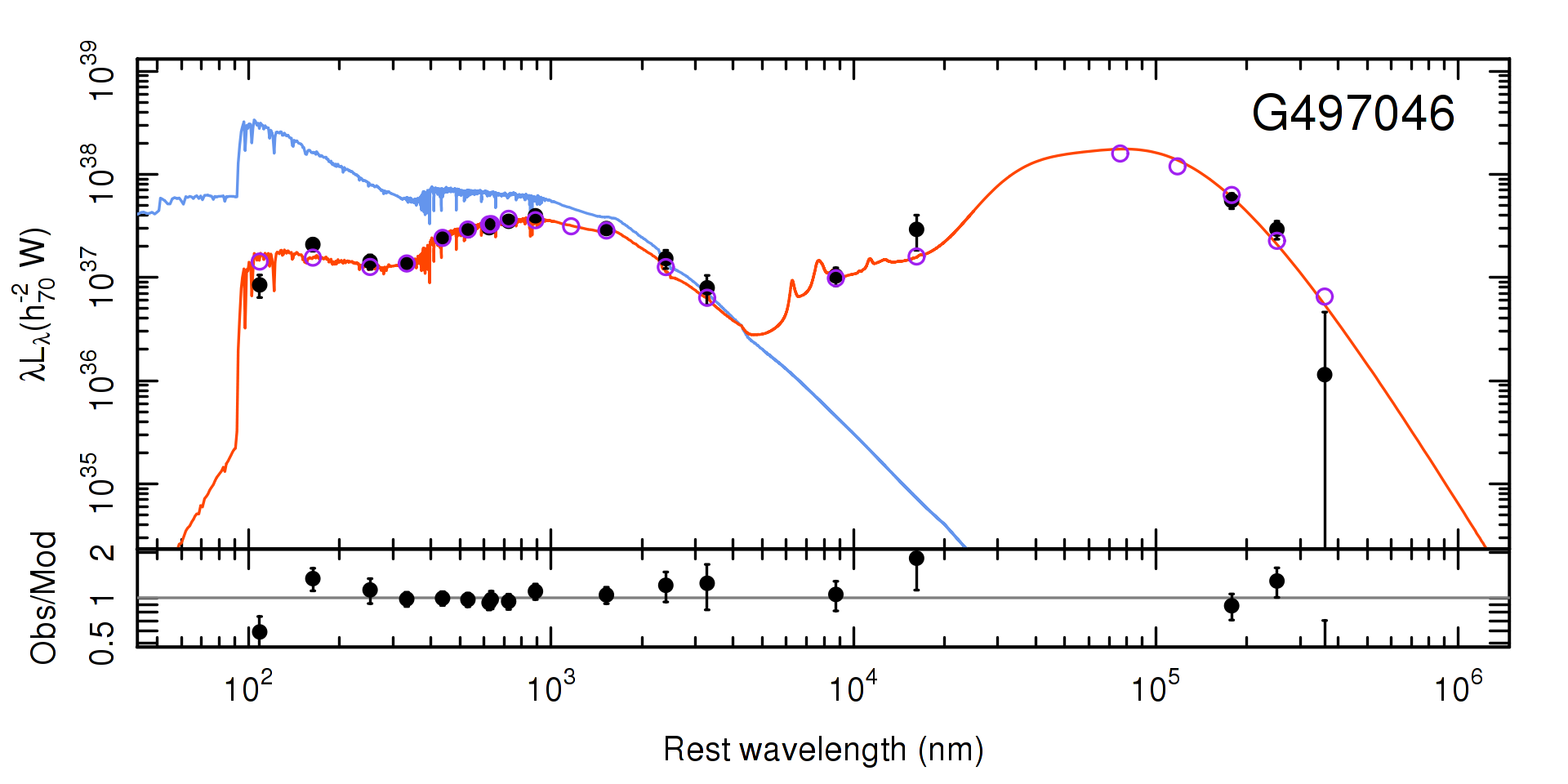}}
\hspace{0.2cm}\subfloat[KiDS~1433]{\hspace{0.cm}\includegraphics[trim={0cm 0cm 0cm 
0cm},clip,width=0.5\textwidth,clip=true]{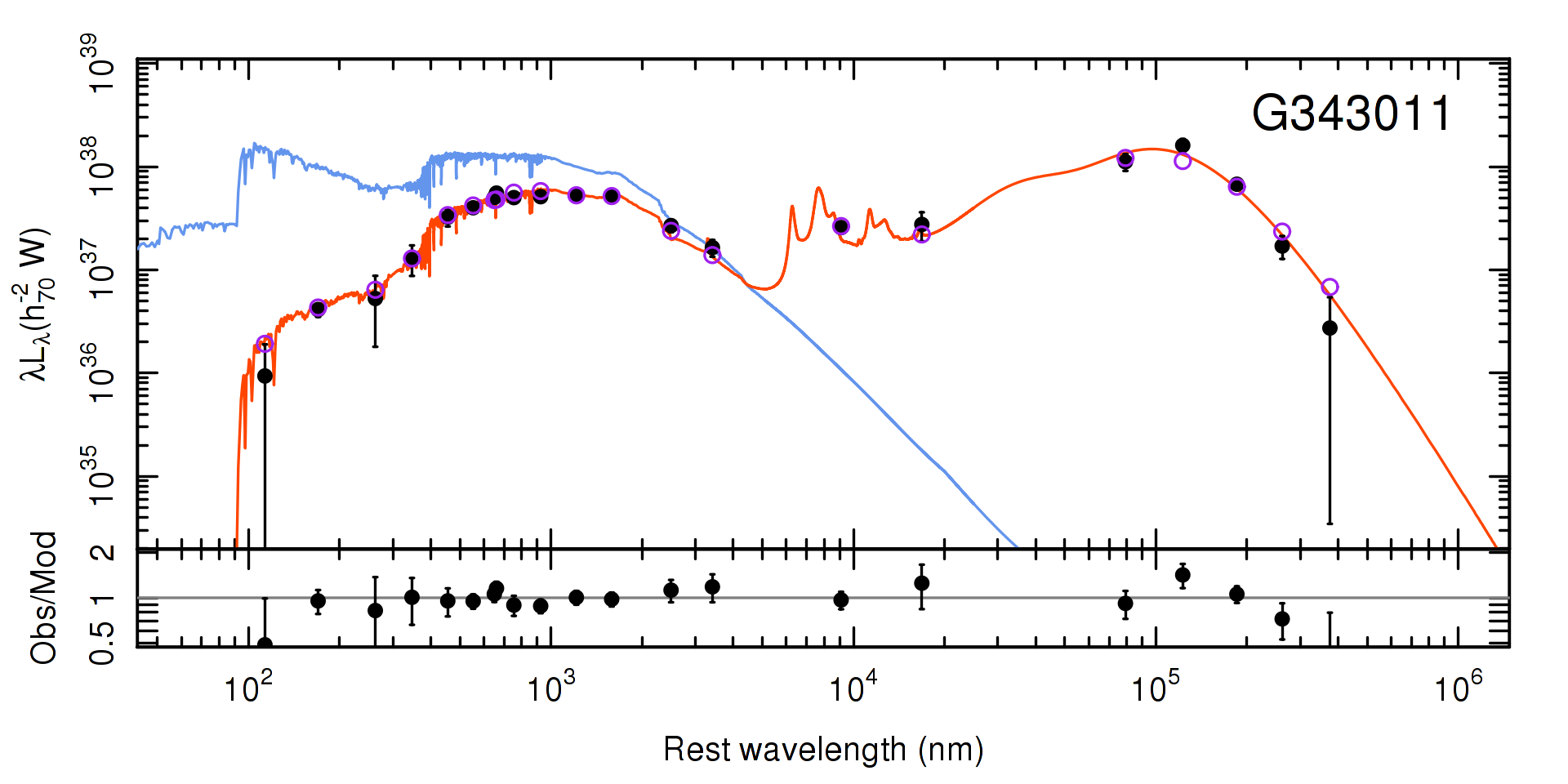}}
\caption{Ultraviolet to far-infrared SEDs of the three BCGs of this work taken from GAMA DR3 \citep{Driver2018}. Dust-attenuated (red curves) and dust-unattenuated (blue curves) MAGPHYS fits are overplotted. Filled black dots and open circles are the observed and model photometry, respectively. The bottom panels show the residuals.}\label{fig:BCG_SEDs}
\end{figure}

\begin{table*}[]\centering
\begin{adjustwidth}{-1cm}{-1cm}
{\small
\begin{center}
\begin{tabular}{ccccccccccc}
\hline\hline
Galaxy  & R.A. & Dec. & $z_{spec}$ &  $\log(M_\star/M_\odot)$ & $\log(L_{\rm dust}/L_\odot)$ & $\log(M_{\rm dust}/M_\odot)$ & SFR & sSFR & sSFR$_{\rm MS}$ & $\log(M_{200}/M_{\odot})$  \\
   & (hh:mm:ss.s) & (dd:mm:ss.s) &  &   & &  & ($M_\odot$/yr) &  (Gyr$^{-1}$) & (Gyr$^{-1}$) & \\ 
  (1) & (2) & (3) & (4) & (5) & (6) & (7) & (8) & (9) & (10) &(11)  \\
  \hline
 KiDS~0837 & 08:37:26.18 &  01:56:39.35 & 0.3958 & $11.64^{+0.06}_{-0.09}$ & $11.56^{+0.06}_{-0.08}$ & $9.43^{+0.09}_{-0.10}$ & $15.2^{+4.4}_{-4.4}$ & $0.04^{+0.02}_{-0.01}$ & 0.08 & $14.29^{+0.12}_{-0.16}$ \\ 
 KiDS~1415 & 14:15:34.71 & -01:08:05.78 & 0.4098 & $11.02^{+0.01}_{-0.15}$ & $11.44^{+0.30}_{-0.07}$ & $8.59^{+0.23}_{-0.24}$ & $23.6^{+0.3}_{-8.0}$ & $0.24^{+0.06}_{-0.05}$ & 0.11 & $14.02^{+0.12}_{-0.21}$ \\ 
 KiDS~1433 & 14:33:29.13 &  02:07:23.36 & 0.3546 &  $11.27^{+0.10}_{-0.16}$ &  $11.62^{+0.05}_{-0.04}$ &  $8.53^{+0.09}_{-0.08}$ &  $21.7^{+5.6}_{-3.0}$ &  $0.12^{+0.07}_{-0.03}$ & 0.07 & $13.74^{+0.12}_{-0.16}$ \\ 
  \hline
\end{tabular}
 \end{center} }
\end{adjustwidth}
\caption{Properties of our targets. (1) BCG name, (2-3) J2000 equatorial coordinates, (4) spectroscopic redshift from GAMA, (5) stellar mass, (6-7) dust luminosity and mass, (8) SFR, (9) specific SFR, (10) specific SFR at the MS estimated using the \citet{Speagle2014} prescription, and (11) richness-based $M_{200}$ mass estimate of the cluster. Columns (5-9) are from MAGPHYS SED fits \citep{Driver2018}.}
\label{tab:BCG_properties}
\end{table*}


\section{Three star-forming BCGs at $z\sim0.4$}\label{sec:BCGsample}
\subsection{Selection of the BCGs in KiDS}\label{sec:BCGselection}
One of the {main} goals of this work is to search for molecular gas reservoirs in distant BCGs and to characterize their stellar and star formation properties. To do this, similarly as C22a, we started by considering a parent sample of 1,484 BCGs with spectroscopic redshifts in the equatorial KiDS-N area. {Of these,  684 BCGs are at $z>0.3$.}  These BCGs were selected by \citet{Radovich2020} using the Data Release (DR)~3 of KiDS \citep{deJong2017}. 
As we are interested in actively star-forming and distant BCGs, we further limited ourselves to the BCGs with clear counterparts in the all-sky data release of the Wide-field Infrared Survey Explorer \citep[WISE;][]{Wright2010} at 22~$\mu$m  in the observer frame (i.e., the W4 channel). We found only six spectroscopically confirmed BCGs at $z>0.3$ with {WISE W4} S/N$>3.3$. The three of these with the highest values, S/N$\sim4.1-6.5$, {were analyzed in C22a}, while in this study, we consider the remaining three, which have WISE W4 S/N$\sim3.4-3.8$. The chosen redshift threshold of $z>0.3$ allows us to focus on distant BCGs and on their evolution, while the cold gas content of more local BCGs, including those of Abell clusters, has been previously and extensively studied \citep[e.g.,][]{Edge2001,Salome_Combes2003,Olivares2019,Rose2023}.

{The three BCGs of this work are} KiDS~0837 (i.e., G345423 at $z=0.3958$), KiDS~1415 (G497046 at $z=0.4098$), and KiDS~1433 (G343011 at $z=0.3546$). In parentheses, we report the ID numbers of the sources from GAMA DR3 \citep{Baldry2018} and the corresponding spectroscopic redshifts.\footnote{\href{http://www.gama-survey.org/dr3/}{http://www.gama-survey.org/dr3/}} The BCGs belong to {the clusters J083727.12+015642, J141534.56-010806, and J143329.76+020719 \citep{Lesci2022},}  which were detected with an S/N = 5.8, 4.0, and 4.2 \citep{Maturi2019,Bellagamba2019}, while they have an intrinsic richness $\lambda_\star=36\pm6$, $28\pm6$, and $17\pm3$, respectively \citep{Maturi2019}. These values correspond to richness-based cluster masses in the range $\log(M_{200}/M_\odot)=13.7-14.3$ (see Table~\ref{tab:BCG_properties}), which are similar to those of the clusters associated with the three BCGs studied in C22a.

These properties on average imply a low probability of $\lesssim1\%$ of the three clusters being false positives \citep[see Fig.~12 of][]{Maturi2019}. Furthermore, we searched by coordinates for our clusters in the literature using the NASA/IPAC Extragalactic Database (NED). 
Both KiDS~0837 and KiDS~1433 BCGs are associated with clusters that are detected in the optical using independent datasets, as outlined in the following.
KiDS~0837 matches the GMBCG~J129.35906+01.94428 \citep{Hao2010} and HSCS~J083731+015837 \citep{Oguri2018} clusters. KiDS~1433 BCG matches the compact group SDSSCGB~22511 \citep{McConnachie2009}, as the two have a projected separation of just 3.9~arcsec (i.e., 19~kpc).

\subsection{Stellar and star formation properties}\label{sec:SFR_Mstar_prop}
In the following, we investigate the multiwavelength properties of the three BCGs of this work. 
In Figure~\ref{fig:BCG_images} we show the color-composite  images of the three galaxies.{
KiDS~1433 is bulge dominated, while KiDS~0837 and KiDS~1415 
appear to be have a more disturbed morphology and clumpy substructures. The visual inspection of the images suggests that our targets {are in} dense environments. Similarly to the KiDS BCGs studied in
C22a, the three BCGs have at least one companion within a projected separation of $\sim4.2$~arcsec (i.e., $\sim$23~kpc at $z=0.4$).}


\subsubsection{Spectral energy distributions}\label{sec:SEDs}
In Figure~\ref{fig:BCG_SEDs} we show the ultraviolet to far-infrared spectral energy distributions (SEDs) of the three sources, taken from the GAMA database.
The SED modeling reported in the panels of the figure corresponds to the MAGPHYS  \citep[][]{daCunha2008} fits by \citet{Driver2018}, {which are publicly available in GAMA DR3}. Photometric data include GALEX \citep{Martin2005,Morrissey2007} in the ultraviolet, SDSS  \citep{York2000} in {the} optical, and the VISTA Kilo-degree Infrared Galaxy Survey 
\citep[VIKING;][]{Edge2013}, WISE \citep{Wright2010}, and {\it Herschel}-ATLAS \citep{Eales2010,Valiante2016} in the near- to far-infrared.

In Table~\ref{tab:BCG_properties} we summarize the main properties of the BCGs, including their stellar, star formation, and dust properties {derived} from the SED fits {\citep{Driver2018}}. The stellar masses of all three BCGs exceed $10^{11}~M_\odot$, which confirms that they are very massive {galaxies}. 

Furthermore, all three BCGs have far-infrared emission from {\it Herschel}, well modeled by a dust component. This supports the scenario that the observed infrared emission is mainly due to star formation, with minimum AGN contamination. Dust masses and luminosities are in the range $\log(M_{\rm dust}/M_\odot)\simeq8.6-9.4$ and $\log(L_{\rm dust}/L_\odot)\simeq11.4-11.6$, respectively, which is typical of luminous infrared galaxies (LIRGs). The SFR estimates based on the ultraviolet to far-infrared SED modeling are high and lie in the range $\sim(15-24)~M_\odot$/yr. These are consistent with but higher than the SFRs of main-sequence (MS) galaxies of similar {masses} and redshifts \citep{Speagle2014}. The only exception is KiDS~0837, which has the lowest SFR, a factor of 2 lower than the average SFR of MS galaxies at its redshift. 

Interestingly, the stellar masses, dust luminosities, and SFRs of the three BCGs of our sample are similar to those of the three star-forming $z\sim0.4$ KiDS BCGs we studied recently in C22a. The six sources were selected {from the KiDS DR3 sample of 684 BCGs with spectroscopic redshifts} $z>0.3$ (Sect.~\ref{sec:BCGselection}). They represent 0.8\% of {this} parent sample and are among the most star-forming BCGs in KiDS. As already pointed out in
C22a, this rare population of BCGs experiences stellar mass assembly and is thus caught in a special phase of their evolution  \citep[see, e.g.,][for similar BCGs]{Castignani2020a,Castignani2020b}.  These BCGs are thus the intermediate-$z$ counterparts of local star-forming BCGs ($>40~M_\odot$/yr) such as the famous Perseus~A and Cygnus~A \citep{FraserMcKelvie2014}.

\subsubsection{Line diagnostics}\label{sec:line_diagnostics}

We now investigate the star formation properties of the BCGs using the spectra found in the GAMA database. We performed a similar analysis to that presented in C22a, as outlined in the following. The SFR is directly proportional to both H$\alpha$ line and [O~II] forbidden-line doublet (3726~\AA, 3729~\AA) luminosities $L_{\rm H\alpha}$ and $L_{\rm [O~II]}$, respectively \citep{Kennicutt1998}. We therefore used these relations to estimate the SFR, calibrated to a \citet{Chabrier2003} IMF.  We also corrected for dust attenuation using the empirical relation by \citet{Garn_Best2010}, which relates dust attenuation {($A_{\rm H\alpha}$)} to stellar mass as follows: 

\begin{equation}
\label{eq:AHa}
 A_{\rm H\alpha} = 0.91 + 0.77 M + 0.11 M^2 - 0.09 M^3\;,
\end{equation}
where $M = \log(M_\star/M_\odot)-10$. With these conventions, {the H$\alpha$- and [O~II]-based SFRs} can be expressed as follows \citep[see, e.g.,][]{Gilbank2010,Zeimann2013,Old2020}:

\begin{equation}
\label{eq:SFR_Ha}
{\frac{{\rm SFR}_{\rm H\alpha}}{M_\odot/{\rm yr}} = 5.0\times10^{-42}\times \frac{L_{\rm H\alpha}}{{\rm erg~s}^{-1}}\times10^{0.4\;A_{\rm H\alpha}}\;,}
\end{equation}

\begin{equation}
\label{eq:SFR_OII}
 \frac{{\rm SFR}_{\rm [O~II]}}{M_\odot/{\rm yr}} =  5.0\times10^{-42}\times\frac{L_{\rm [O~II]}}{{\rm erg~s}^{-1}}\times\frac{10^{0.4\;A_{\rm H\alpha}}}{r_{\rm lines}}\;,
\end{equation}
where $r_{\rm lines}$ is the ratio  of extinguished [O~II] to H$\alpha$ fluxes.  

\begin{figure*}[ht!]\centering
\captionsetup[subfigure]{labelformat=empty}
\subfloat[]{\hspace{-0.7cm}\includegraphics[trim={1cm 2cm 3.5cm 
5cm},clip,width=0.5\textwidth,clip=true]{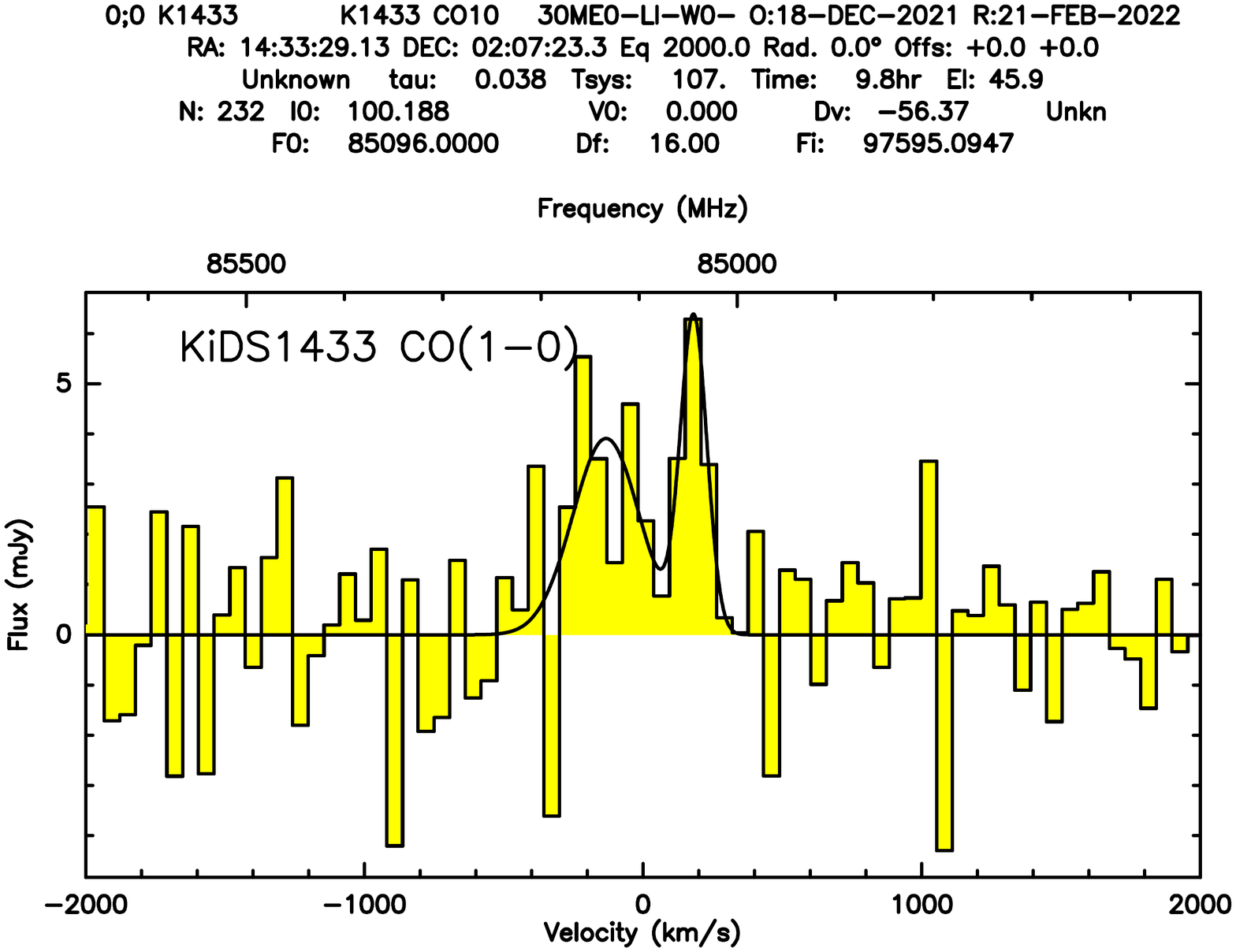}}
\subfloat[]{\hspace{0.7cm}\includegraphics[trim={1cm 2cm 3.5cm 
5cm},clip,width=0.5\textwidth,clip=true]{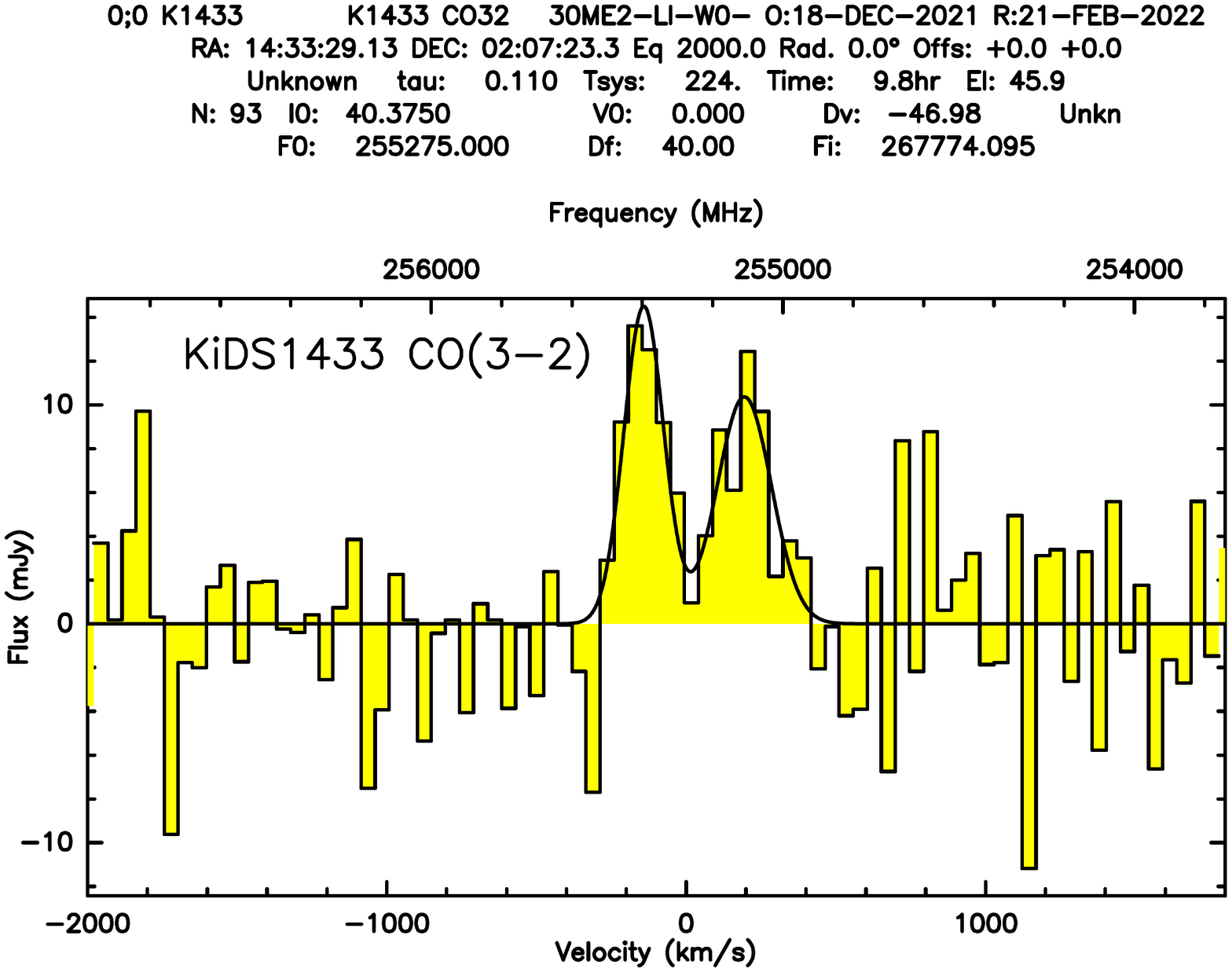}}\\
\caption{Baseline-subtracted CO(1$\rightarrow$0) and CO(3$\rightarrow$2) spectra of KiDS~1433. Solid lines show the double Gaussian fits to the emission lines. See text for details. The flux (y-axis) is plotted against the relative velocity with respect to the BCG redshift, as in Table~\ref{tab:BCG_properties} (bottom x-axis) and the observer frame frequency (top x-axis).}\label{fig:BCG_spectra}
\end{figure*}

We found GAMA spectra for all three BCGs considered in this work, taken with the AAOmega/2dF multifiber spectrograph on the 3.9~m Anglo-Australian Telescope. Gaussian fits to several lines are also available \citep{Gordon2017}. Unfortunately, H$\alpha$ is redshifted to wavelengths longer than 8890~\AA\ for all three sources, and they are therefore beyond the  wavelength range of $3750~\AA\lesssim\lambda\lesssim8850~\AA$ that is covered by the spectrograph \citep{Hopkins2013}. 

However, the [O~II] doublet is detected in the GAMA spectra of all three BCGs. Following \citet{Gilbank2010}, we summed the fluxes of the two components of the [O~II] doublet for each sources, adding their uncertainties in quadrature. We refer to this sum simply as [O~II], as in Eq.~\ref{eq:SFR_OII}. The resulting [O~II] doublet fluxes of KiDS~0837, 1415, and 1433 are $F_{\rm [O~II]}=(14.8\pm4.7)$, $(75.7\pm25.6)$, and $(104.2\pm31.6)$, respectively, in units of  $10^{-17}$~erg~s$^{-1}$~cm$^{-2}$. 


With these fluxes, Eq.~\ref{eq:SFR_OII} yields an ${\rm SFR}_{\rm [O~II]}=(5.5\pm1.7)$, $(21.9\pm7.4)$, and $(25.1\pm7.6)~M_\odot$/yr  for KiDS~0837, 1415 and 1433, respectively, where we assumed $r_{\rm lines}=0.5$, following \citet{Gilbank2010}. With {the only}  possible exception of KiDS~0837, which has the lowest SFR, we find a good agreement within the reported uncertainties when we compare the [O~II]-based SFRs with those {obtained by SED fits (Table~\ref{tab:BCG_properties}), which} are used throughout this work. This is remarkable considering that discrepancies, even up to a factor of $\sim3$, are common when different  estimates of the SFR are compared \citep[e.g.,][]{Calzetti2013}. 

\subsubsection{WISE infrared colors}\label{sec:IR_diagnostics}

Finally, we investigated the star formation properties and the AGN contamination in the infrared by inspecting the position of our sources in the infrared W1-W2 vs. W2-W3 color-color WISE diagram \citep{Jarrett2017}. {In this diagram, sources with W1-W2$\gtrsim$0.8~mag can be considered to be AGN, while lower values of W1-W2 are associated with other types of sources. The W1-W2 colors of the three KiDS BCGs of this work lie in the range between (0.36-0.55)~mag and are therefore not classified as AGN. On the basis of their WISE colors,} both KiDS~1415 and 1433 are indeed classified as starbursts, {while KiDS~0837 falls in the region of galaxies with SFRs below the starburst level.} However, the last BCG has only a 3$\sigma$ lower limit of W3$>11.2$~mag. This implies W2-W3$<2.7$, so that the spheroid class is still possible for the source. Overall, these findings strengthen the selection of our three targets as star-forming galaxies and suggest that any possible contamination from a circum-nuclear dusty torus at infrared wavelengths is likely negligible, which is also supported by the generally good agreement between the SED- and line-based SFRs. Interestingly, C22a found similar line-based SFRs and infrared colors  for the additional three star-forming intermediate-$z$ KiDS BCGs that come from our selection. These results thus show that the six BCGs together belong to a homogeneous population of BCGs in terms of stellar, star formation, and dust properties. However, as we discuss in Sect.~\ref{sec:thoughts_gas_BCGs}, their molecular properties are more heterogeneous.

\section{Observations and analysis}\label{sec:observations_and_data_reduction}
\subsection{Millimeter IRAM 30m observations}
We observed the three KiDS BCGs using the IRAM 30m telescope at Pico Veleta in Spain. The observations were carried out on December 15-21, 2021 (ID: 176-21; PI: G.~Castignani). We used the Eight Mixer Receiver (EMIR) to observe CO(J$\rightarrow$J-1) emission lines from the BCGs, where J is a positive integer denoting the total angular momentum. For each galaxy, the specific CO(J$\rightarrow$J-1) transitions were chosen to maximize the detection likelihood in terms of the ratio of the predicted signal to the expected root mean square (rms) noise. 

The E090, E150, and E230 receivers, operating at $\sim$1-3~mm, offer 4$\times$4~GHz instantaneous bandwidth covered by the correlators. For the KiDS~0837 and KiDS~1415 BCGs, we used the two E090 and E150 receivers simultaneously to observe CO(1$\rightarrow$0) and CO(2$\rightarrow$1), redshifted to 3~mm and 2~mm in the observer frame, respectively. For all three targets, we also used the E090 and E230 receivers to simultaneously target the CO(1$\rightarrow$0) and CO(3$\rightarrow$2) lines, redshifted to 3~mm and 1~mm, respectively. We note that we did not observe  KiDS~1433 in CO(2$\rightarrow$1) as the line is redshifted to 
170.189~GHz in the observer frame, which is too close to the 183~GHz water vapor line. The line is thus prohibited because of low atmospheric transmission. 

The IRAM 30m half-power beam width (HPBW) is $\sim16$~arcsec~$\frac{\lambda_{\rm obs}}{2~{\rm mm}}$ \citep{Kramer2013}, where $\lambda_{\rm obs}$ is the observer frame wavelength. All targets are thus unresolved by our observations, assuming a CO-to-optical size ratio $\sim0.5$ \citep{Young1995}; see Fig.~\ref{fig:BCG_images}. The wobbler-switching mode was used for all the observations to minimize the impact of atmospheric variability.
The Wideband Line Multiple Autocorrelator (WILMA) was used to cover the 4$\times$4~GHz bandwidth in each linear polarization.   We also simultaneously recorded the data with the fast Fourier Transform Spectrometers (FTS) as a backup at 200 kHz resolution.

We encountered variable weather conditions during our observations. Overall, we had average system temperatures typical of {the winter} season and equal to $T_{\rm sys}\simeq107$-$116$~K, 228-278~K, and 183-224~K at 3, 2, and 1~mm,
respectively. As calibrators, we used the quasars
PKS~1253-055 and OJ~287 for pointing, and PKS~1253-055 for the focus.
We summarize our observations in Table~\ref{tab:BCG_properties_mol_gas}.  The data reduction and analysis were performed using the {\sc CLASS} software of the {\sc GILDAS}  package\footnote{https://www.iram.fr/IRAMFR/GILDAS/}. 
Our results are presented in Sect.~\ref{sec:results}, and in the following, we describe our analysis of the IRAM-30m spectra for the three targeted BCGs.

\begin{table*}[tb]\centering
\begin{adjustwidth}{-1.1cm}{}
\begin{center}
\begin{tabular}{cccccccccccccc}
\hline
\hline
  &   &    &   &  &  &  &  &  & \hspace{-0.2cm}{\small integration} \\
 Galaxy  &  $z_{spec}$ &   CO(J$\rightarrow$J-1)  &  $\nu_{\rm obs}$ & $S_{\rm CO(J\rightarrow J-1)}\,\Delta\varv$   & S/N & FWHM & $L^\prime_{\rm CO(J\rightarrow J-1)}$ & $z_{\rm CO(J\rightarrow J-1)}$ & time \\ 
   &  & & (GHz) &  (Jy~km~s$^{-1}$) & & (km~s$^{-1}$) & ($10^{10}$~K~km~s$^{-1}$~pc$^2$) & & (hr)  \\ 
 (1) & (2) & (3) & (4) & (5) & (6) & (7) & (8) & (9) & (10) \\
 \hline
KiDS~0837 & 0.3958 & $1\rightarrow0$ & 82.584 & $<0.83$ & --- & --- & $<0.67$ & --- & 2.3  \\
  &  & $2\rightarrow1$ & 165.165 & $<2.64$ & --- & --- & $<0.53$ & --- & 1.0   \\
  &  & $3\rightarrow2$ & 247.740 & $<2.25$ & --- & --- & $<0.20$ & --- & 1.3  \\
  \hline
KiDS~1415 & 0.4098 &  $1\rightarrow0$ & 81.764 & $<0.62$ & --- & --- & $<0.54$ & --- & 3.5    \\
  &  & $2\rightarrow1$ & 163.525 &   $<1.56$ & --- & --- & $<0.34$ & --- & 1.6 \\
  &  & $3\rightarrow2$ & 245.280 & $<1.69$ & --- & --- & $<0.16$ & --- & 1.9 \\
  \hline
KiDS~1433 & 0.3546 & $1\rightarrow0$ & 85.096 & $2.10\pm0.42$ (Total) & 5.0 & $551\pm111$  & $1.35\pm0.27$ & $0.3546\pm0.0002$ &  2.5 \\
 &  &  & & $1.18\pm0.34$ ($\varv<0$) & 3.5 & $283\pm98$  & $0.76\pm0.22$ & $0.3541\pm0.0001$ &  \\
 &  &  & & $0.74\pm0.22$ ($\varv>0$) & 3.4 & $112\pm33$  & $0.48\pm0.14$ &  $0.35521\pm0.00005$ &  \\
 &  &\\ 
  &  & $3\rightarrow2$ & 255.275 &  $4.66\pm0.84$ (Total) & 5.5 & $497\pm77$ & $0.332\pm0.060$  & $0.3547\pm0.0002$ &  2.5 \\ 
 &  &  & & $2.42\pm0.51$ ($\varv<0$) & 4.7 & $157\pm33$  & $0.173\pm0.037$ & $0.35413\pm0.00006$ &  \\
 &  &  & & $2.36\pm0.63$ ($\varv>0$) & 3.7 & $214\pm67$  & $0.168\pm0.045$ & $0.35525\pm0.00009$ &  \\
%
%

\hline
\end{tabular}
\end{center}
\caption{Results of our CO observations. Column descriptions: (1) galaxy name,  (2) spectroscopic redshift as in Table~\ref{tab:BCG_properties}, (3-4) CO(J$\rightarrow$J-1) transition and observer frame frequency, (5) CO(J$\rightarrow$J-1) velocity-integrated flux, (6) S/N of the CO(J$\rightarrow$J-1) detection, (7) FWHM of the CO(J$\rightarrow$J-1) line; (8) CO(J$\rightarrow$J-1) velocity-integrated luminosity, (9) redshift derived from the CO(J$\rightarrow$J-1) line, and (10) on-source integration time (double polar). Absent values are denoted with the long dash. For  KiDS~0837 and KiDS~1415, the reported upper limits are at 3$\sigma$, {estimated at a resolution of 300~km~s$^{-1}$.}}
\label{tab:BCG_properties_mol_gas}
\end{adjustwidth}
\end{table*}

\begin{table*}[]
\begin{center}
\begin{tabular}{ccccccccccc}
\hline\hline
Galaxy & $z_{\rm spec}$ & excitation ratio & $M_{H_2}$ & $\tau_{\rm dep}$ & $\frac{M_{H_2}}{M_\star}$ & $\frac{M_{H_2}}{M_{\rm dust}}$ & $\tau_{\rm dep, MS}$ & $\big(\frac{M_{H_2}}{M_\star}\big)_{\rm MS}$  \\
 & & & ($10^{10}M_\odot$) &  (Gyr) &  & & (Gyr) \\
 (1) & (2) & (3) & (4) & (5) & (6) & (7) & (8) & (9) \\
 \hline
 KiDS~0837 & 0.3958 & --- & $<2.9$ & $<1.9$ &  $<0.07$ & $<11$ & $1.4^{+0.3}_{-0.2}$ & $0.07^{+0.12}_{-0.05}$\\
  KiDS~1415 & 0.4098 & --- & $<2.4$ & $<1.0$ & $<0.23$  & $<61$ & $1.3^{+0.2}_{-0.2}$ & $0.13^{+0.21}_{-0.08}$\\
KiDS~1433 & 0.3546 & $r_{31}=0.25\pm0.07$ & $5.9\pm1.2$ & $2.7^{+0.7}_{-0.9}$ & $0.32^{+0.12}_{-0.10}$ & $174^{+45}_{-53}$ & $1.4^{+0.2}_{-0.2}$ & $0.09^{+0.16}_{-0.06}$ \\
\hline
\end{tabular}
\end{center}
 \caption{Molecular gas properties of the BCGs. Column descriptions: (1) BCG name, (2) spectroscopic redshift as in Table~\ref{tab:BCG_properties}, (3) excitation ratio, (4) molecular gas mass $M_{H_2} =\alpha_{\rm CO} L^\prime_{\rm CO(J\rightarrow J-1)}$, where $\alpha_{\rm CO}=4.36~M_\odot\,({\rm K~km~s}^{-1}~{\rm pc}^2)^{-1}$ is assumed, (5) depletion timescale  $\tau_{\rm dep}=M_{H_2}/{\rm SFR}$; (6) molecular gas-to-stellar mass ratio, (7) molecular gas-to-dust mass ratio, and (8-9) depletion timescale and molecular gas-to-stellar mass ratio predicted for MS field galaxies with the redshift and stellar mass of our targets, following the prescription by \citet{Tacconi2018}, calibrated to $\alpha_{\rm CO}=4.36~M_\odot\,({\rm K~km~s}^{-1}~{\rm pc}^2)^{-1}$. For  KiDS~0837 and KiDS~1415, the reported upper limits are at 3$\sigma$.} 
\label{tab:COresults}
\end{table*}

%

\subsection{Molecular gas properties}\label{sec:IRAM30m_analysis}
For KiDS~1433, our observations yield clear detections of CO in the first and third transitions at an $\textrm{S/N}\simeq5.0-5.5$. The corresponding spectra are shown in Fig.~\ref{fig:BCG_spectra}.
We fit the total emission using a single Gaussian for each of the CO(1$\rightarrow$0) and CO(3$\rightarrow$2)  spectra of KiDS~1433. As the CO(1$\rightarrow$0) and CO(3$\rightarrow$2) lines of KiDS~1433 are double-peaked,  we additionally fit the total emission for each spectrum using a combination of  two Gaussian curves that are overlaid on the spectra in Fig.~\ref{fig:BCG_spectra}. In Table~\ref{tab:BCG_properties_mol_gas} we report the best-fit results for the single-Gaussian fits (denoted {\it Total}) and for the double-Gaussian fits (labeled $\varv>0$ and $\varv<0$). For the last ones, $\varv>0$ ($\varv<0$) denotes the fit to the CO emission that peaks at positive (negative) velocity with respect to the BCG redshift. As outlined in Table~\ref{tab:BCG_properties}, we observe an overall agreement between the single- and double-Gaussian fits in terms of velocity-integrated flux, full width at half maximum (FWHM), and S/N. Our CO-based redshift measurement also agrees well with the BCG redshift that is estimated with optical spectral lines.


The observed double-peaked CO emission is probably likely due to the KiDS~1433 BCG alone, and we further discuss the underlying structural parameters in Sect.~\ref{sec:double_horn_modeling}. However, the BCG has two nearby companions at a projected separation of $\sim$4.2-4.7~arcsec {(see Fig.~\ref{fig:BCG_images}).} They have high AMICO cluster membership probabilities of $0.93-0.96$ and are therefore photometrically selected cluster members. These BCG companions may contribute to the CO(1$\rightarrow$0) and CO(3$\rightarrow$2) emission, which may accordingly result in the observed double-horn profile.  Nevertheless, this appears to be unlikely as the efficiency is highly suppressed for CO(3$\rightarrow$2), down to 54\% at most, given the angular separations of the BCG companions. For CO(1$\rightarrow$0), the efficiency suppression is only marginal, at a level of 93$\%$. However, it is worth noting that the positive and negative peaks have similar height, and each of them similarly contributes to $(46\pm11)\%$ of the total emission for both CO(1$\rightarrow$0) and CO(3$\rightarrow$2). Similarly, the $\varv>0$ peaks and the $\varv<0$ peaks have similar velocities in the two panels of Fig.~\ref{fig:BCG_spectra}. These results indicate a scenario in which the double-peak emission in the CO(1$\rightarrow$0) and CO(3$\rightarrow$2) spectra has the same physical origin, and we interpret it as CO emission from the BCG KiDS~1433 itself. However, higher angular resolution observations, for example, with ALMA or NOEMA interferometers, are needed to spatially resolve the molecular gas reservoir and firmly test the proposed scenario.

For the other two targeted BCGs, KiDS~0837 and 1415, we did not detect any CO emission in their spectra. We therefore assigned upper limits to the CO line fluxes as follows. We first removed the baseline in each spectrum with a linear fit. We then estimated the rms noise level for the antenna temperature (Ta$^\ast$). Using  the standard conversions outlined below, we converted the rms into a 3$\sigma$ upper limit to the integrated CO flux at a resolution of 300~km~s$^{-1}$, which is the typical FWHM for massive galaxies and BCGs in particular \citep{Edge2001,Dunne2021}.

The results of our analysis are reported in Table~\ref{tab:BCG_properties_mol_gas} for all three BCGs we targeted in this work. We applied standard efficiency corrections to convert i) the antenna temperature Ta$^\ast$ into the main beam temperature $T_{\rm mb}$, and then ii) $T_{\rm mb}$ into the corresponding CO line flux, with a conversion factor of 5~Jy/K. In particular, we adopted the following efficiency  corrections: $T_{\rm mb}/T{\rm a}^\ast = 1.2$, $1.4$ and $2.0$ for $\sim$3, 2, and 1~mm observations, respectively.\footnote{https://www.iram.es/IRAMES/mainWiki/Iram30mEfficiencies} 

We then derived the CO(J$\rightarrow$J-1) luminosity $L^{\prime}_{\rm CO(J\rightarrow J-1)}$, in units {of} K~km~s$^{-1}$~pc$^2$, from the velocity-integrated CO(J$\rightarrow$J-1) flux $S_{\rm CO(J\rightarrow J-1)}\,\Delta\varv\ $, in units {of} Jy~km~s$^{-1}$.
To do this, we used Eq.~(3) from \citet{Solomon_VandenBout2005},
\begin{equation}
\label{eq:LpCO}
 L^{\prime}_{\rm CO(J\rightarrow J-1)}=3.25\times10^7\,S_{\rm CO(J\rightarrow J-1)}\,\Delta\varv\,\nu_{\rm obs}^{-2}\,D_L^2\,(1+z)^{-3}\,,
\end{equation}
where $\nu_{\rm obs}$ is the observer frame frequency in GHz of the CO(J$\rightarrow$J-1) transition, $D_L$ is the luminosity distance in Mpc, and $z$ is the redshift of the BCG.

We then converted the velocity-integrated CO(1$\rightarrow$0) line fluxes into total $H_2$ gas masses, or into their upper limits. The CO(1$\rightarrow$0) transition is preferred to higher-J transitions as it does not require any assumption on the excitation ratio
$r_{\rm J1}= L^{\prime}_{\rm CO(J\rightarrow J-1)}/L^{\prime}_{\rm CO(1\rightarrow0)}$.

The three KiDS BCGs have similar sSFRs within the range  ${\rm sSFR}\simeq(0.5-2.2)\times {\rm sSFR}_{\rm MS}$ (see Table~\ref{tab:BCG_properties}). They are all lower than the value of $3\times {\rm sSFR}_{\rm MS}$, below which the galaxy sSFR is within the characteristic scatter of the MS. To estimate $H_2$ gas masses, we therefore assumed a Galactic CO-to-$H_2$ conversion factor of $\alpha_{\rm CO}=4.36~M_\odot\,({\rm K~km~s}^{-1}~{\rm pc}^2)^{-1}$, which is commonly used for star-forming galaxies at the MS. The use of a single $\alpha_{\rm CO}$ conversion factor also allowed us to perform a homogeneous comparison in terms of gas content (see Sect.~\ref{sec:results}).



Last, we used the $H_2$ gas mass estimates of the BCGs to estimate a number of related quantities, or upper limits.  The first quantity is the depletion timescale $\tau_{\rm dep}=M_{H_2}/{\rm SFR}$, which is the characteristic time in which the molecular gas reservoirs are depleted. To compute $\tau_{\rm dep}$ , we used the SED-based SFR estimates reported in Table~\ref{tab:BCG_properties}. 
Similarly, we  also estimated the molecular-gas-to-stellar-mass ratio, $M_{H_2}/M_\star$ and the ratio of the molecular gas to the dust masses.  In addition, as a comparison, we computed the depletion time $\tau_{\rm dep, MS}$ and the molecular gas-to-stellar mass ratio $\big(\frac{M_{H_2}}{M_\star}\big)_{\rm MS}$ for MS galaxies in the field with a redshift and stellar mass equal to those of our BCGs, by using the empirical prescriptions by \citet{Tacconi2018}, calibrated to the CO-to-$H_2$ conversion factor of $\alpha_{\rm CO}=4.36~M_\odot\,({\rm K~km~s}^{-1}~{\rm pc}^2)^{-1}$ we used here. 

The resulting molecular gas properties for the three KiDS BCGs are summarized in Table~\ref{tab:COresults}. As KiDS~1433 BCG is detected in both CO(1$\rightarrow$0) and CO(3$\rightarrow$2), we also report the associated $r_{31}$ excitation ratio in the table.

\begin{figure*}[h!]\centering
\subfloat{\hspace{0.2cm}\includegraphics[trim={1.5cm 0.cm 2.5cm 
0.5cm},clip,width=0.5\textwidth,clip=true]{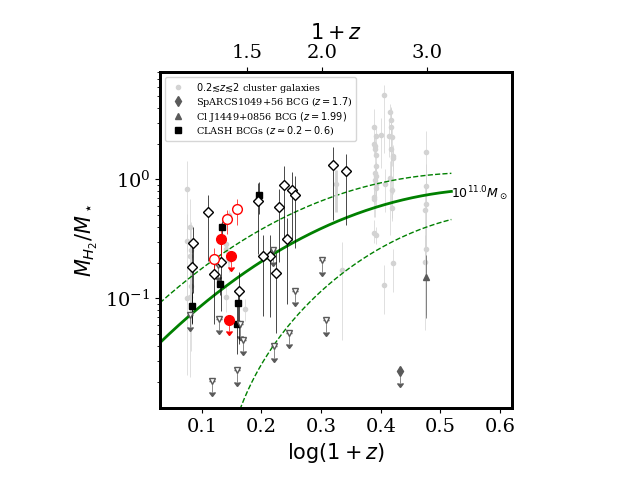}}
\subfloat{\hspace{0.2cm}\includegraphics[trim={1.5cm 0.cm 2.5cm 
0.5cm},clip,width=0.5\textwidth,clip=true]{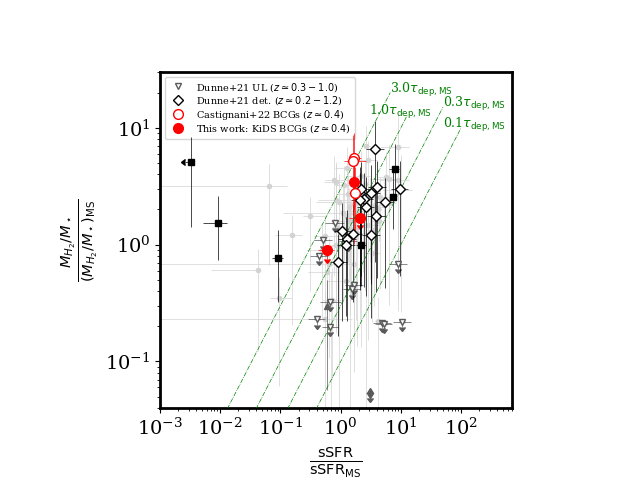}}\\
\caption{{ Molecular gas properties of distant  BCGs and cluster galaxies observed in CO}. Left: Evolution of the molecular gas-to-stellar mass ratio as a function of the redshift for cluster galaxies at $0.2\lesssim z\lesssim2$ observed in CO.  
The solid green curve is the scaling relation found  by \citet{Tacconi2018} for field galaxies in the MS and  with $\log(M_\star/M_\odot)$=11, which corresponds to the median stellar mass of all sources in the plot. The dashed green lines show the statistical 1$\sigma$ uncertainties in the model. Right: Molecular gas-to-stellar mass ratio plotted against the sSFR for the cluster galaxies, normalized to the corresponding MS values using the relations for the ratio and the SFR by \citet{Tacconi2018} and \citet{Speagle2014}, respectively. The dot-dashed green lines show different depletion times in units of the depletion time at the MS. The points in both panels show cluster galaxies with molecular gas observations. The color code is reported at the top left of each panel. KiDS BCGs of this work are reported as filled red circles, and open circles refer to KiDS BCGs in C22a. Additional distant BCGs from the literature are also highlighted. We refer to the text for further details.}\label{fig:mol_gas1}
\end{figure*}

\begin{figure}[]\centering
\subfloat{\hspace{0.2cm}\includegraphics[trim={1.5cm 0.cm 2.5cm 
1.5cm},clip,width=0.4\textwidth,clip=true]{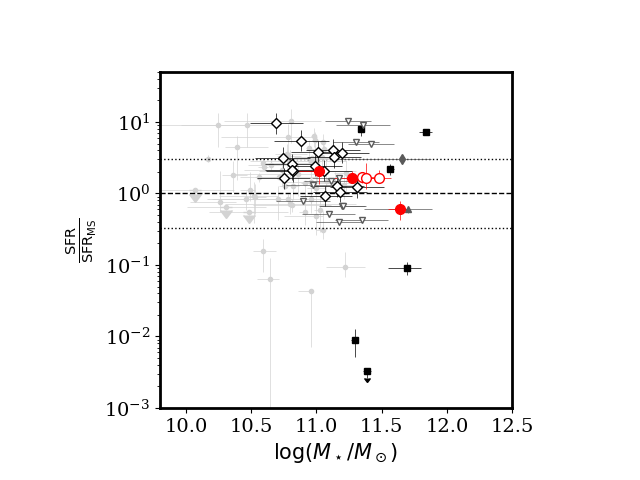}}\\
\subfloat{\hspace{0.2cm}\includegraphics[trim={1.5cm 0.cm 2.5cm 
1.5cm},clip,width=0.4\textwidth,clip=true]{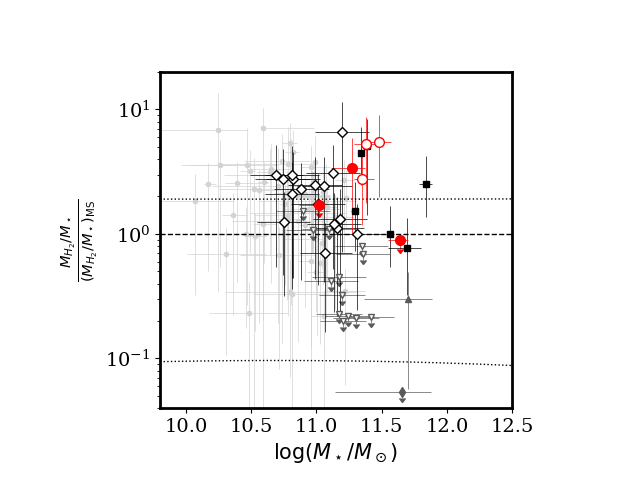}}\\
\subfloat{\hspace{0.2cm}\includegraphics[trim={1.5cm 0.cm 2.5cm 
1.5cm},clip,width=0.4\textwidth,clip=true]{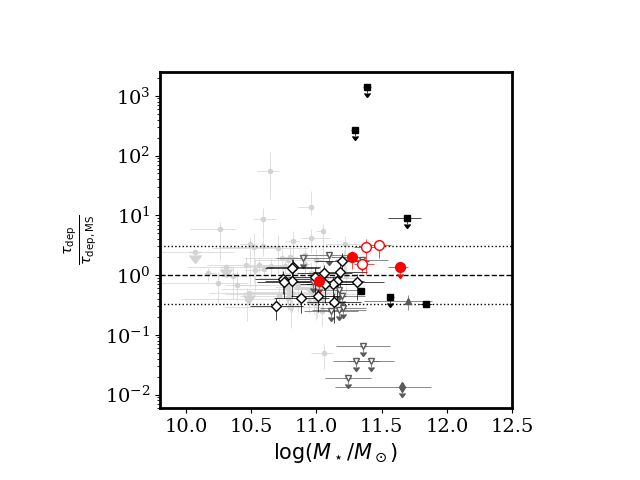}}
\caption{{ SFR (top), molecular gas-to-stellar mass ratio (center), and depletion time (bottom) as a function of the stellar mass for distant BCGs and cluster galaxies observed in CO.} The y-axis values are all normalized to the corresponding MS values using the relations by \citet{Speagle2014} and \citet{Tacconi2018}. The horizontal dashed lines correspond to y-axis values equal to unity, and the dotted lines are the fiducial uncertainties associated with the MS. The uncertainty is chosen equal to $\pm0.48$~dex for the top and bottom panels because the MS is commonly identified by $1/3<{\rm SFR}/{\rm SFR}_{\rm MS}<3$. For the central panel, the plotted uncertainties are estimated at redshift $z=0.5$. The color-coding for the data points is the same as in Fig.~\ref{fig:mol_gas1}.}\label{fig:mol_gas3}
\end{figure}

\begin{figure*}[th!]\centering
\captionsetup[subfigure]{labelformat=empty}
\subfloat[]{\hspace{0.cm}\includegraphics[trim={2cm 0cm 3.2cm 
1.5cm},clip,width=0.4\textwidth,clip=true]{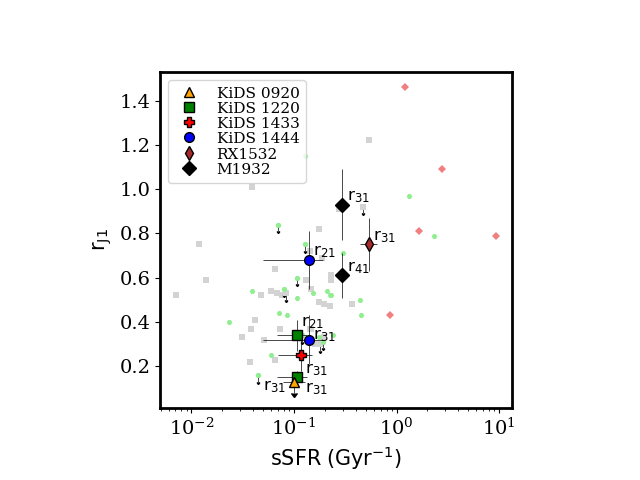}}
\subfloat[]{\hspace{0.cm}\includegraphics[trim={2cm 0cm 3.2cm 
1.5cm},clip,width=0.4\textwidth,clip=true]{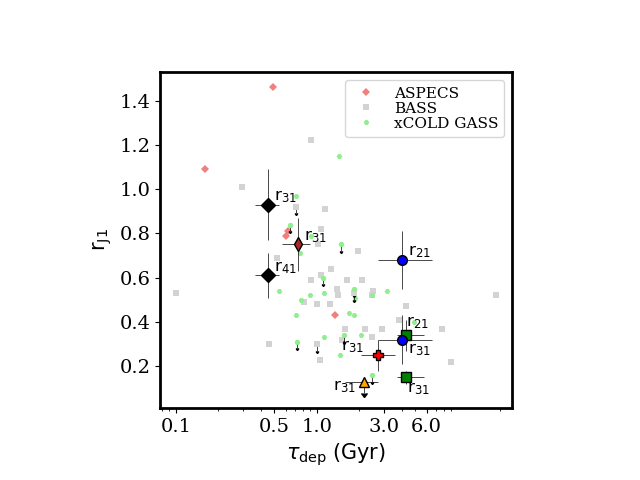}}
\caption{Excitation ratios $r_{J1}$ plotted against the sSFR (left) and the depletion time (right) for intermediate-redshift BCGs with multiple CO(J$\rightarrow$J-1) detections. {The color-code for the BCGs and the comparison sources (ASPECS, BASS, and XCOLD GASS) is reported at the top of the two panels.}}\label{fig:excitation}
\end{figure*}

\begin{figure}[htb!]\centering
\captionsetup[subfigure]{labelformat=empty}
\subfloat[]{\hspace{0.cm}\includegraphics[trim={0cm 0cm 0cm 
0cm},clip,trim = {2.5cm 0.5cm 1cm 1.2cm}, width=0.36\textwidth,clip=true]{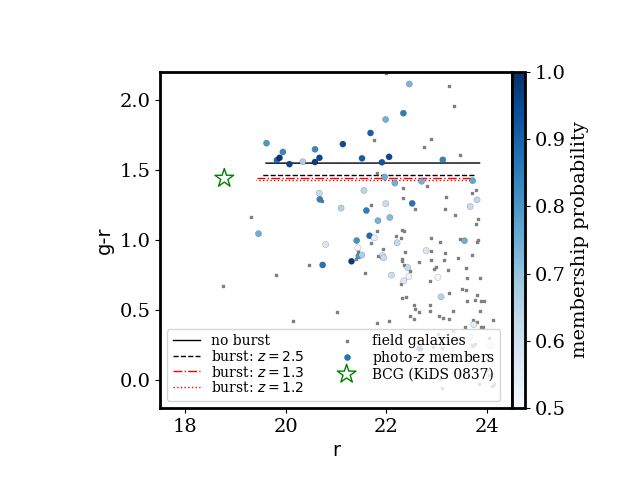}}
\hspace{0.1cm}\subfloat[]{\hspace{0.cm}\includegraphics[trim={0cm 0cm 0cm 0cm},clip,trim = {2.5cm 0.5cm 1cm 1.2cm},width=0.36\textwidth,clip=true]{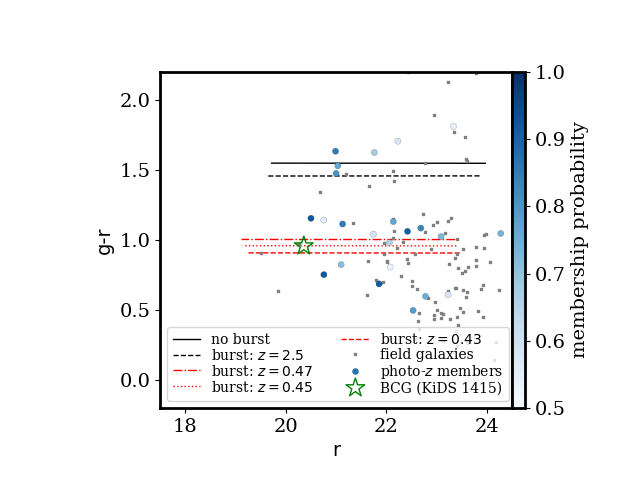}}
\hspace{0.1cm}\subfloat[]{\hspace{0.cm}\includegraphics[trim={0cm 0cm 0cm 
0cm},clip,trim = {2.5cm 0.5cm 1cm 1.2cm},width=0.36\textwidth,clip=true]{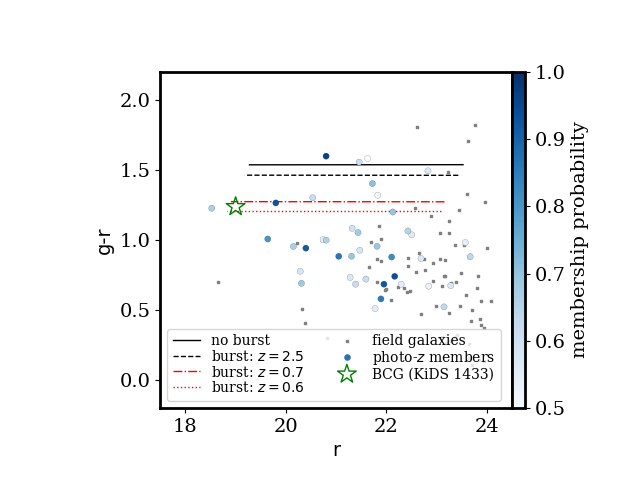}}
\caption{Color-magnitude plots for the sources in the field of the BCGs KiDS~0837 (top), KiDS~1415 (center), and KiDS~1433 (bottom). Field galaxies (gray points), photometrically selected cluster members (blue points), {color-coded according to their cluster membership probabilities,} and the BCGs (green stars) are highlighted. Red-sequence models are shown as horizontal lines. See the legend and text for further details.}\label{fig:CM_plots}
\end{figure}

\section{Results}\label{sec:results}

One of the main results of this work is that with the new CO(1$\rightarrow$0) and CO(3$\rightarrow$2) detections of KiDS~1433, we notably increase the still limited sample of {distant BCGs} with clear detections in two distinct CO lines. To the best of our knowledge, only five additional BCGs are detected in multiple CO transitions, all of them at intermediate redshifts $z\sim0.4$. They are the star-forming BCGs KiDS~0920, 1220, and 1444 {(C22a)}, and the BCGs RX1532 \citep{Castignani2020a} and M1932 \citet{Fogarty2019} from the CLASH survey \citep{Postman2012}. In the following, we present and further discuss our results in detail.

\subsection{Comparison sample}\label{sec:comparison_sample}
We compared the molecular gas properties of our targeted BCGs with those of existing samples of cluster galaxies with molecular gas observations. To do this, we considered the sample of $z<2$ cluster galaxies with $M_\star>10^{10}~M_\odot$ described in Sect.~4.1 of C22a and references therein, which is an update of the compilation reported in Tables~A.1 and A.2 of \citet{Castignani2020b}, to which we refer for further details. The only difference with respect to the comparison sample used in C22a is as follows. In C22a, we mostly considered cluster galaxies with detections in CO. However, for two out of the three BCGs of this work we are only able to set upper limits to their molecular gas mass. Therefore, we now considered the full sample of \citet{Dunne2021}, consisting of distant BCGs within the SpARCS survey. More specifically, we included the BCGs in their sample in our analysis with CO detections and $3\sigma$ upper limits.
The full list of comparison galaxies we considered comprises 97 cluster galaxies over 55 clusters at $z\sim0.2-2$ and stellar masses in the range $\log(M_\star/M_\odot)\simeq10-12$.
They include 40 BCGs that we briefly outline in the following. 



We included the gas-rich ClJ1449+0856 BCG at $z = 1.99$ \citep{Strazzullo2018}, with $M_{H_2}=7.5\times10^{10}~M_\odot$ \citep{Coogan2018}, {and} the SpARCS~104922.6+564032.5 BCG at $z=1.7$ \citep{Webb2017,Barfety2022}, with an upper limit of $M_{H_2}<1.1\times10^{10}~M_\odot$ to the molecular gas mass.
The recent molecular gas  survey by \citet{Dunne2021} with ALMA adds another 30 SpARCS BCGs: seventeen of them are  detected in CO(2$\rightarrow$1) and S/N$>3$, and we report for the remaining 13 $3\sigma$ upper limits to their $H_2$ gas mass here.  These BCGs span the redshift range $z\sim0.2-1.2$ and their molecular gas masses lie in the range  $M_{H_2}\sim(10^{10}-10^{11})~M_\odot$, or upper limits  in the range $M_{H_2}\lesssim(0.3-4.8)\times10^{10}~M_\odot$.

We included six CLASH BCGs at intermediate redshifts $z\sim0.2-0.6$, similarly to our KiDS targets, as outlined in the following. RX1532 and M1932  were clearly detected in CO(1-0) at S/N$>6$ with the IRAM 30m \citep{Castignani2020a} and  ALMA \citep{Fogarty2019}, respectively. They are gas-rich systems, with $M_{H_2}\sim10^{11}~M_\odot$. We also included the CLASH BCGs M0329, A1423, M1206, and M2129, which were detected at S/N$\gtrsim3$ in CO(1$\rightarrow$0) or CO(2$\rightarrow$1) as part of our IRAM 30m campaign \citep{Castignani2020a}, and their molecular gas masses lie in the range $M_{H_2}\sim(10^{10}-10^{11})~M_\odot$.

Last, we included the three BCGs KiDS~0920, 1220, and 1444 at $z\sim0.4$, with gas masses in the range $M_{H_2}\sim(5-15)\times10^{10}~M_\odot$, as inferred from CO(1$\rightarrow$0) detections {(C22a)}.
In Sect.~\ref{sec:gas_SFR_tdep} we use this sample of distant cluster galaxies observed in CO, including BCGs, to compare stellar mass, SFR, molecular gas content, and depletion time with the three BCGs of this work.

In addition to the $\sim100$ cluster galaxies with observations in CO listed above, we considered a second compilation of field galaxies with molecular gas observations in the first and third CO transitions that we use in Sect.~\ref{sec:excitation_ratios} to investigate the excitation ratios. Analogously to  what we discussed in Sect~4.3 of C22a, this comparison sample of sources with observations in multiple CO transitions includes 5 distant ($2.5<z<2.7$) galaxies that are part of the ALMA Spectroscopic Survey in the {\it Hubble} Ultra Deep Field
\citep[ASPECS;][]{Boogaard2020}, and 61 local galaxies and AGN from \citet{Lamperti2020}, which are drawn from the xCOLD GASS survey \citep{Saintonge2017} and the BAT AGN Spectroscopic Survey (BASS)\footnote{http://www.bass-survey.com/}.

\subsection{Gas fraction, star formation, and depletion time}\label{sec:gas_SFR_tdep}

In this section, we place the molecular gas, stellar, and  star formation properties of our targeted BCGs in a broader context. To do this, we used the compilation of $\sim100$ cluster galaxies with $M_\star$ and SFR estimates as well as observations of their molecular gas, as outlined in Sect.~\ref{sec:comparison_sample}. 

Figure~\ref{fig:mol_gas1} (left panel) shows the ratio of molecular gas to stellar mass  $M_{H_2}/M_\star$ as a function of redshift. In the right panel of Fig.~\ref{fig:mol_gas1}, we instead show the $M_{H_2}/M_\star$ ratio as a function of the 
specific SFR, both normalized to the MS. Figure~\ref{fig:mol_gas3} displays the SFR, $M_{H_2}/M_\star$, and the depletion time ($\tau_{\rm dep})$, all normalized to their MS values, as a function of the stellar mass.

In both figures, we report data of distant cluster galaxies with molecular gas observations. They include the three targeted BCGs of this work (filled red circles), the three KiDS BCGs of our recent C22a work (open red circles), and the compilation of $\sim100$ cluster galaxies of our comparison sample, described in Sect.~\ref{sec:comparison_sample}. In the latter, BCGs from the literature are highlighted, and the remaining cluster galaxies in our comparison sample are shown in background (gray dots). To allow a homogeneous comparison, we used $\alpha_{\rm CO}=4.36~M_\odot\,({\rm K~km~s}^{-1}~{\rm pc}^2)^{-1}$ for all data points and the overplotted \citet{Tacconi2018} relations in both figures.

Our targeted BCGs have high SED-based stellar masses in the range of $\log(M_\star/M_\odot)\sim(11.0-11.6)$; see Table~\ref{tab:BCG_properties}. As illustrated in Fig.~\ref{fig:mol_gas3}, {these BCGs fall in the upper tail of the $M_\star$ distribution for cluster galaxies, which is populated by BCGs in our comparison sample.}  These results place the targeted sources of this work among the most massive cluster galaxies at intermediate redshifts, strengthening their selection as  BCGs \citep{Radovich2020}, which we further discuss in Sect.~\ref{sec:color_magnitude_plots}.

Furthermore, Figs.~\ref{fig:mol_gas1} and \ref{fig:mol_gas3} both show that KiDS~1433 is very rich in molecular gas having $M_{H_2}\simeq6\times10^{10}~M_\odot$ and a $H_2$-to-stellar mass ratio of $M_{H_2}/M_\star\simeq0.32$ (see also Table~\ref{tab:COresults}). These values agree with those we found in 
C22a for similarly star-forming $z\sim0.4$ BCGs KiDS~0920, 1220, and 1444, suggesting that the four BCGs together belong to the same population of star-forming and gas-rich BCGs. Interestingly, the other two $z\sim0.4$ BCGs that we targeted in this work, that is, KiDS~0837 and 1415, have similar star formation levels as the other four KiDS BCGs, well within the MS; see Fig~\ref{fig:mol_gas1} (right panel) and Fig.~\ref{fig:mol_gas3} (top panel). However, KiDS~0837 and 1415 BCGs were not detected in CO with our observations, and we place upper limits of $M_{H_2}/M_\star\lesssim0.07$ and 0.23, respectively, which are consistent with the most stringent upper limits found for distant BCGs in previous studies \citep[e.g.,][]{Castignani2020a,Castignani2020b,Castignani2020c,Dunne2021}. For a {direct} comparison, we also refer to the upper limits from the literature that are reported in Fig.~\ref{fig:mol_gas1}.

{Under the assumption that the total infrared emission is a tracer of ongoing star formation fueled by the molecular gas reservoir \citep[e.g.,][]{Solomon1997,Tacconi2018},} a possible interpretation is that our millimeter observations are just {below} the CO detection limit for KiDS~0837 and 1415 BCGs. Their total infrared fluxes are 2.5 and $1.8\times10^{12}$~erg~cm$^{-2}$, respectively, as inferred from the total SED-based dust luminosities reported in Table~\ref{tab:BCG_properties}. Conversely, the remaining four KiDS BCGs detected in CO all have {slightly} higher infrared fluxes, in the range of $\sim(2.6-4.7)\times10^{12}$~erg~cm$^{-2}$. However, based on our CO upper limits, even with deeper observations (e.g., with NOEMA or with the ALMA interferometers), we expect KiDS~0837 and 1415 to have an $H_2$ amount at the level of the MS at most, which is  significantly lower than that we infer for the four KiDS BCGs with CO {detections}, whose $H_2$ gas masses lie well above the MS levels (Fig.~\ref{fig:mol_gas3}, central panel).

\subsection{Star-forming BCGs in KiDS. Some thoughts on gas-rich versus gas-poor sources.}\label{sec:thoughts_gas_BCGs}

To summarize, the six KiDS BCGs considered in this and our previous work {(C22a)} have similar levels of star formation activity within the MS. However, as illustrated in Fig.~\ref{fig:mol_gas3}, those with CO detections tend to have larger $H_2$ gas reservoirs and longer depletion times ($\tau_{\rm dep}$), normalized to the MS values, than those found in the KiDS BCGs with only upper limits in CO. Similarly, there is also a dichotomy in terms of the $H_2$-to-dust-mass ratio. While the four KiDS BCGs with CO detections have high ratios $M_{H_2}/M_{\rm dust}\simeq(170-300)$, the two BCGs with only upper limits in CO have much lower ratios, well below the value $M_{H_2}/M_{\rm dust}\simeq100$, which is typical of distant star-forming galaxies \citep{Scoville2014,Scoville2016,Berta2016}.

{To explain the high values of $\tau_{\rm dep}$, $M_{H_2}/M_\star$, and   $M_{H_2}/M_{\rm dust}$ observed in the KiDS BCGs with CO detections, including {KiDS~1433} that is the subject of this work, we suggest that they are observed in a peculiar and rare phase of the BCG evolution, in which a substantial amount of the $H_2$ gas has recently been accreted, but still not efficiently converted into stars. As has been suggested in 
C22a, the accretion may have occurred   via interaction with the cluster companions.  The other star-forming BCGs (KiDS~0837 and 1415), with only upper limits in CO, may be at a different and possibly later stage of the BCG evolution, where the $H_2$ gas reservoir has been exhausted or is undergoing rapid consumption at a timescale $\tau_{\rm dep}\lesssim(1-2)$~Gyr, while star formation is still ongoing. These results suggest that molecular gas depletion precedes star formation quenching in intermediate-$z$ and star-forming BCGs. This result agrees with what \citet{Castignani2022c} found locally for early-type galaxies in the field of {the} Virgo cluster.}

{Furthermore, KiDS~0837 and KiDS~1415 BCGs appear to have clumpy substructures (Fig.~\ref{fig:BCG_images}), which is partially at odds with the other KiDS BCGs with detections in CO 
(see also Fig.~1 of C22a). Therefore, it might be that compact components and clumpy morphology favor the rapid exhaustion of molecular gas and ultimately help to quench the BCGs. This scenario agrees with that proposed in \citet{Castignani2020b}, who studied a sample of distant star-forming BCGs mostly with only upper limits in CO and  hence short depletion times. By combining molecular gas observations with a {\it Hubble Space Telescope} morphological analysis, we speculated that compact components and clumpy morphology may favor the rapid exhaustion of molecular gas and ultimately help to quench the BCGs. This scenario might also be applicable to the KiDS~0837 and KiDS~1415 BCGs.}





\subsection{CO-to-$H_2$ conversion and metallicities}
In the previous section, we discussed an environment-driven scenario that may explain the large molecular gas reservoirs that are observed in some distant star-forming KiDS BCGs. However, as already pointed out in C22a, a lower $\alpha_{\rm CO}$ conversion factor than the Galactic one, as is usually used for starbursts, would lead to lower $M_{H_2}$ and $\tau_{\rm dep}$.

{Therefore, we reconsidered the $H_2$ gas masses of all six KiDS BCGs studied in this work and in
C22a. To do this, we  estimated the CO-to-$H_2$ conversion factor using  existing metallicity-dependent prescriptions. In particular, we adopted estimates of the metallicity taken from GAMA DR3, which come as outputs of the SED fits with MAGPHYS by \citet[][see Fig.~\ref{fig:BCG_SEDs}]{Driver2018}.}
{We then used the metallicity values to estimate the CO-to-$H_2$ conversion factor using Eq.~4 of \citet{Tacconi2018}, which corresponds to the geometric mean of the $\alpha_{\rm CO}$ prescriptions reported in \citet{Bolatto2013} and \citet{Genzel2012,Genzel2015}. Similarly to \citet{Tacconi2018}, we relied on the relation $\log Z=12+\log({\rm O/H})$, which is the metallicity on the \citet{Pettini_Pagel2004} scale, and the solar metallicity was assumed to be equal to  $\log Z_\odot=8.67$ \citep[see, e.g.,][for a review]{Asplund2009}.}

{In Table~\ref{tab:alphaCO_metallicity} we report the results of our analysis. The six KiDS BCGs span a broad range of metallicities, between $Z/Z_\odot\sim0.4-1.8$, which reflects a similarly broad range in $\alpha_{\rm CO}\sim2.7-9.3$, while estimates for individual galaxies have non-negligible uncertainties. On average,  $\alpha_{\rm CO}=(4.8\pm2.1)~M_\odot\,({\rm K~km~s}^{-1}~{\rm pc}^2)^{-1}$ holds, where we report the median value and the rms dispersion around the median. These results imply that our choice of adopting a Galactic  $\alpha_{\rm CO}$ is reasonable for our targeted BCGs, and thus the $M_{H_2}$ estimates do not appear to be biased.}

\begin{table}[tb]\centering
\begin{center}
\begin{tabular}{cccccc}
\hline
\hline\\
 Galaxy  &  $Z/Z_\odot$ & $\alpha_{\rm CO}(Z)$ \\ 
     &   & ($M_\odot\,({\rm K~km~s}^{-1}~{\rm pc}^2)^{-1}$) \\ 
 (1) & (2) & (3) \smallskip\\
 \hline\\
KiDS~0920 & $1.82^{+0.15}_{-0.23}$ & $2.69^{+0.16}_{-0.25}$ \smallskip \\
KiDS~1220 & $0.77^{+0.39}_{-0.27}$ & $5.32^{+2.34}_{-1.62}$ \smallskip \\
KiDS~1444 & $1.01^{+0.36}_{-0.31}$ & $4.24^{+1.23}_{-1.06}$ \smallskip \\
\hline\\
KiDS~0837 & $0.81^{+0.33}_{-0.18}$ & $5.10^{+1.78}_{-0.97}$ \smallskip \\
KiDS~1415 & $0.96^{+0.25}_{-0.74}$ & $4.42^{+0.95}_{-2.80}$ \smallskip \\
KiDS~1433 & $0.43^{+0.39}_{-0.16}$ & $9.27^{+8.86}_{-3.63}$ \smallskip \\
\hline
\end{tabular}
\end{center}
\caption{{Metallicity-dependent CO-to-$H_2$ conversion factor. Column description: (1) galaxy name,  (2) metallicity derived from GAMA DR3 SED fits with MAGPHYS by \citet{Driver2018}, and (3) $\alpha_{\rm CO}$ conversion factor estimated using metallicity-dependent prescriptions \citep{Bolatto2013,Genzel2012,Genzel2015}; see text for further details.}}
\label{tab:alphaCO_metallicity}
\end{table}


\subsection{Excitation ratios}\label{sec:excitation_ratios}
Motivated by the results presented above, we now further investigate the differences of the targeted BCGs with respect to those studied in previous works by means of the excitation ratio. To the best of our knowledge, there are only a few distant BCGs with CO detections in different transitions, all at intermediate redshifts. These are RX1532 
\citep[$z=0.361$;][]{Castignani2020a} and M1932 \citep[$z=0.353$;][]{Fogarty2019} from the CLASH survey, with high levels of star formation (${\rm SFR}\gtrsim100~M_\odot/{\rm yr}$). In addition, there are other four BCGs from KiDS at redshifts between $z=0.32-0.44$ and with lower levels of star formation (${\rm SFR}\simeq22-34~M_\odot/{\rm yr}$). These are {the} KiDS~0920, 1220, 1444, and 1433 BCGs. The first three were the subject of our recent C22a study, and for the present analysis, we included KiDS~1433, which we targeted in this work.
As the sources are all detected in CO(1$\rightarrow$0) as well as in CO(2$\rightarrow$1),  CO(3$\rightarrow$2), or CO(4$\rightarrow$3), we considered the $r_{21}$,  $r_{32}$, and  $r_{43}$ excitation ratios whenever available. 

Similarly to the analysis presented in C22a, in Figure~\ref{fig:excitation} we show the excitation ratios $r_{\rm J1}$ of the BCGs against the sSFR (left) and the depletion time (right). For the latter, we used $\alpha_{\rm CO}=4.36~M_\odot\,({\rm K~km~s}^{-1}~{\rm pc}^2)^{-1}$ for all BCGs to allow a homogeneous comparison.
KiDS~1433 of this work has similar properties in terms of  excitation ratio, sSFR, and depletion time as those of the similarly star-forming and gas-rich KiDS BCGs considered in
C22a, whose excitation ratios lie in the range $r_{31}\sim0.1\pm0.3$.  On the other hand, the CLASH BCGs (M1932 and RX1532), which have higher levels of star formation activity than {the} KiDS BCGs, also have more excited gas reservoirs, with higher values of $r_{31}\sim0.8-0.9$, as well as lower depletion times and higher sSFR values.

As all these six BCGs, including KiDS~1433 from this work, have observations in both CO(1$\rightarrow$0) and CO(3$\rightarrow$2), we include in Fig.~\ref{fig:excitation} the ASPECS, BASS, and xCOLD GASS galaxies introduced in Sect.~\ref{sec:comparison_sample} for a comparison, similarly to what we did in C22a. When we consider the $r_{31}$ values of the BCGs (KiDS and CLASH) and the comparison galaxies together, we find that the excitation ratio is {correlated} with sSFR and anticorrelated with the depletion time at a level of 3.2$\sigma$ (${\rm p-value}=1.62\times10^{-3}$) and 4.8$\sigma$~ (${\rm p-value}=1.26\times10^{-6}$), as found with a Spearman test, respectively.
A correlation at the level of 2.8$\sigma$~ (${\rm p-value}=4.8\times10^{-3}$) is also found when considering the KiDS and CLASH BCGs alone in the $r_{31}$ vs. sSFR plane. These significances are {similar as but also slightly higher} than those found by a similar analysis in C22a, where KiDS~1433 was not included.

These results strengthen the scenario proposed in our previous studies ( \citet{Castignani2020a,Castignani2022a}) such that while similarly large amounts of molecular gas reservoirs, with $M_{H_2}\simeq(0.5-1.4)\times10^{11}~M_\odot$ and $M_{H_2}/M_\star\simeq0.1-0.6$, are often found in star-forming (SFR$\gtrsim20~M_\odot$/yr) BCGs at intermediate-$z$, only the most star-forming (SFR$\gtrsim100~M_\odot$/yr) and cool-core BCGs such as RX1532 and M1932 are able to excite the gas at the highest levels, for which the cool-core cluster environments likely
favor the condensation and the inflow of gas onto the BCGs
themselves \citep{Castignani2020a}. {Low excitation values $r_{\rm J1}\lesssim0.4$ are instead found for the BCGs with high $\tau_{\rm dep}\gtrsim2$~Gyr such as KiDS~1220 and 1433, which have double-horn CO spectra that are indicative of a low gas concentration  (see also Sect.~\ref{sec:double_horn_modeling}). These results are reasonable, as in the regime of low star formation efficiency, that is, high $\tau_{\rm dep}$,  a large amount of gas might be located far out in the disk and  not involved in the star formation, as is the case for the nearby giant elliptical Centaurus~A \citep{Salome2017}, for example.}

{Alternatively,} we recall that the IRAM 30m HPBW is equal to $\sim$24 and 10~arcsec for the CO(1$\rightarrow$0) and CO(3$\rightarrow$2) transitions, respectively. Similarly to the three KiDS BCGs of 
C22a, KiDS~1433 of this work has nearby companions within 10~arcsec in projection. These are clearly visible in Fig.~\ref{fig:BCG_images} and might contribute to the observed CO(1$\rightarrow$0) emission, while the contamination is likely negligible for CO(3$\rightarrow$2), as we further discussed in Sect.~\ref{sec:IRAM30m_analysis}. 
The relatively low $r_{\rm 31}$ values of the KiDS BCGs might therefore be explained when the BCG companions are gas rich. High-resolution interferometric observations are needed for a conclusive answer.

\subsection{Color-magnitude plots}\label{sec:color_magnitude_plots}

In this section, we investigate the cluster galaxy population of the three KiDS clusters of this work  by means of color-magnitude plots, with a particular focus on the BCGs. We performed an analysis similar to that presented in C22a for the additional three KiDS BCGs with molecular gas observations. We started by selecting sources in the KiDS DR3 within a separation of 2~arcmin from each of the three targeted BCGs.

In Fig.~\ref{fig:CM_plots} we report the corresponding \textsf{g-r} versus \textsf{r} plots, where \textsf{g} and \textsf{r} are AB apparent magnitudes, obtained with the OmegaCAM camera on the VST (see Sect.~\ref{sec:BCGselection}). At the redshifts of our three targeted BCGs, the rest frame 4000~\AA~ break is redshifted to $\sim(5500-5600)$~\AA\  between the \textsf{g} and \textsf{r} filters, whose  effective wavelengths are 4800~\AA~ and 6250~\AA, respectively. The chosen filters are thus optimal to detect  the red sequence in the considered clusters. In the color-magnitude plots, the BCGs are denoted with green stars. We also distinguish field galaxies (gray points) from photometric cluster members {(blue points), which were selected as galaxies with cluster membership probabilities greater than 50\%, as was further discussed} in Sect~5.4 of C22a. Using both GAMA and SDSS spectroscopic databases, we did not find any other spectroscopic cluster member in addition to the BCG themselves.

Visual inspection of the color-magnitude plots shows that our KiDS BCGs are the brightest of the selected cluster members. The one exception is a photometrically selected member with \textsf{r}=18.5 and a projected separation of 70~arcsec ($\sim$0.4~Mpc) in projection from the KiDS~1433 BCG. {The galaxy has an AMICO cluster membership probability} of $0.67$ and is only $\sim0.5$~mag brighter than {the} BCG, but with a photometric redshift of $0.21\pm0.06$. It is therefore likely a foreground source.

{For the KiDS~0837 cluster, we detect a strong clustering of red sources at \textsf{g-r}$\sim1.5$.} This suggests that the KiDS~0837 cluster core is sufficiently mature and its galaxy population is well evolved in terms of color properties. {For the other two clusters, the region of the plot around \textsf{g-r}$\sim1.5$ is instead only weakly populated by cluster members. However,} { cluster members with bluer colors}  \textsf{g-r}$\sim1.0-1.3$, {similar to those of the associated BCGs appear to lie there as well,}  which may indicate that ongoing star formation activity is not limited to the BCGs. The BCGs KiDS~0837, 1415, and 1433 are systematically blue, that is, with relatively low values of \textsf{g-r}$=1.4$, $1.0$, and $1.2$, respectively. {These values are indeed $\sim(0.3-0.5)$~mag lower than the median value of \textsf{g-r}$=1.70^{+0.11}_{-0.25}$,\footnote{{Here the uncertainties correspond to the 68\% confidence interval.}} which is associated with $0.3<z<0.5$ BCGs in KiDS DR3 \citep{Radovich2020}. } These results generally agree with the fact that we selected the BCGs to be star-forming so that they are caught in a rare phase of their evolution.

These results encouraged us to perform a red-sequence modeling, similarly to previous studies \citep{Kotyla2016,Castignani2019}. We followed the same approach as was adopted in Sect.~5.3 of
C22a, to which we refer for further details. In particular, we used the tool Galaxy Evolutionary Synthesis Models (GALEV)  \footnote{http://www.galev.org} \citep{Kotulla2009}.  We adopted chemically consistent models of elliptical galaxies with a galaxy formation redshift of $z=8$.  We also assumed that the total initial mass of the galaxy is between $1\times10^{10}~M_\odot$ and $5\times10^{11}~M_\odot$.

With these assumptions, we computed different red-sequence models, which we report  as horizontal lines in Fig.~\ref{fig:CM_plots}, as further outlined in the following. For all three color-magnitude plots, we show a model assuming no burst, which yields the reddest possible red sequence at the cluster redshift, as well as a model with a star formation burst at $z=2.5$, exponentially declining with time. For all three clusters, the colors of the latter model are only slightly bluer than those inferred from models without a star formation burst. While these two models both nicely reproduce the colors of the reddest sources, with colors \textsf{g-r}$\simeq1.5$, they hardly reproduce the bluer colors of BCGs. This is particularly true for the BCGs KiDS~1415 and 1433.

These considerations motivated us to perform additional models with more recent bursts of star formation, which are plotted in Fig.~\ref{fig:CM_plots} as red horizontal lines. More recent bursts of star  formation at $z\sim1.2-1.3$, $z\sim0.45$, and $z\sim0.6-0.7$  are  indeed needed to better reproduce the observed \textsf{g-r} colors and the observed \textsf{r}-band magnitudes of  the three KiDS~0837, 1415, and 1433 BCGs, respectively, as well. 

For the BCG KiDS~0837, a burst at $z=2.5$ still reproduces the BCG \textsf{g-r} color fairly well. However, a later burst at $z\sim1.2-1.3$ appears to be favored as it implies both a bluer color and a brighter magnitude, closer to those of the BCG.  Similarly to the case of KiDS~1444 discussed in 
C22a, we note that the adopted red-sequence models are not able to reach the low magnitude \textsf{r}$\simeq18.8$ of {the} KiDS~0837 BCG. This result suggests that a faster mass growth (e.g., via mergers) than predicted by the closed-box GALEV evolutionary models would be a viable alternative for the BCG.\\

{The color-magnitude plots and red sequence modeling together favor recent bursts of star formation for the BCGs. These results agree well with the fact that the BCGs are star forming. As further discussed in Sect.~\ref{sec:gas_SFR_tdep} and in C22a, recently accreted pristine gas may explain the observed properties of the BCGs.
For KiDS~1415, in particular, our modeling implies a remarkably recent {(300~Myr)} burst of star formation. This result, together with the high star formation activity and high infrared luminosity of the BCG, suggests that the source may be interacting with the nearby companion(s), which are clearly visible in Figure~\ref{fig:BCG_images}.}

\begin{figure*}[]\centering
\captionsetup[subfigure]{labelformat=empty}
\subfloat[(a)]{\hspace{0.cm}\includegraphics[trim={0cm 0cm 1cm 
1cm},clip,width=0.45\textwidth,clip=true]{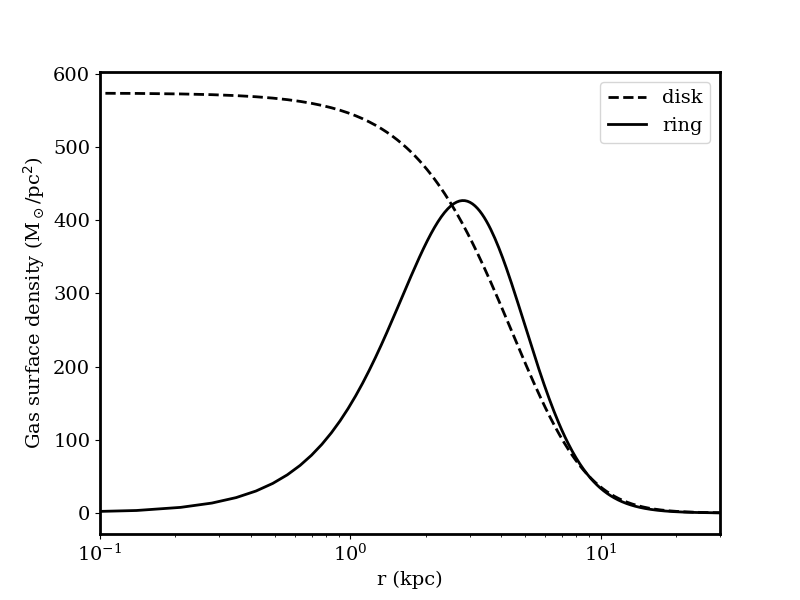}}
\hspace{0.1cm}\subfloat[(b)]{\hspace{0.cm}\includegraphics[trim={0cm 0cm 1cm 1cm},clip,width=0.45\textwidth,clip=true]{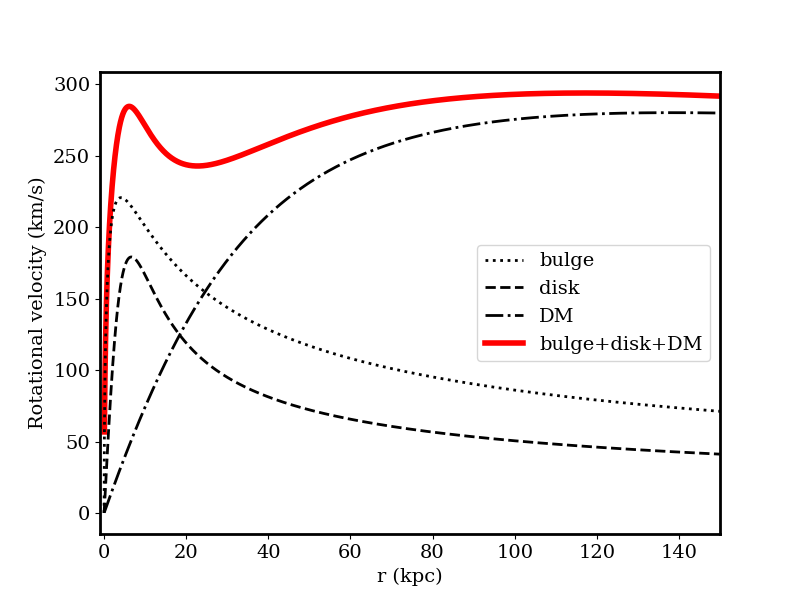}}\\
\hspace{0.1cm}\subfloat[(c)]{\hspace{0.cm}\includegraphics[trim={0cm 0cm 1cm 1cm},clip,width=0.45\textwidth,clip=true]{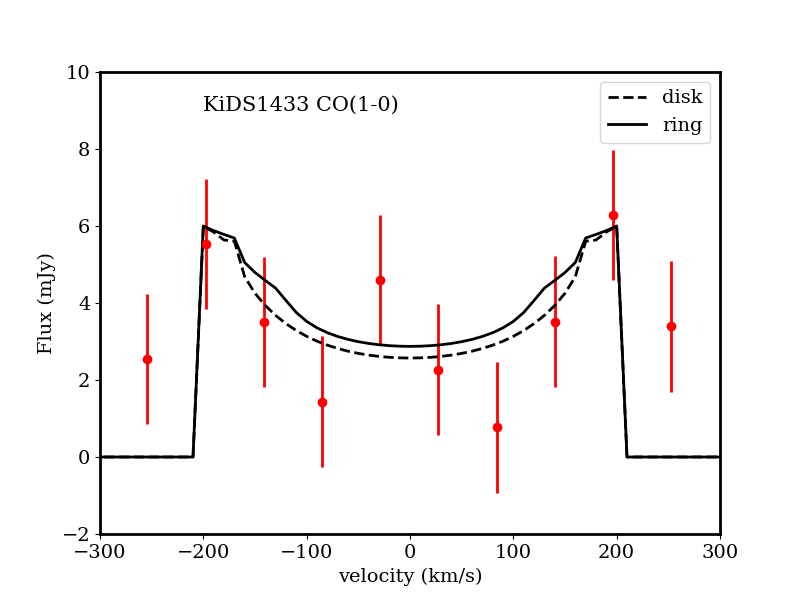}}
\hspace{0.1cm}\subfloat[(d)]{\hspace{0.cm}\includegraphics[trim={0cm 0cm 1cm 1cm},clip,width=0.45\textwidth,clip=true]{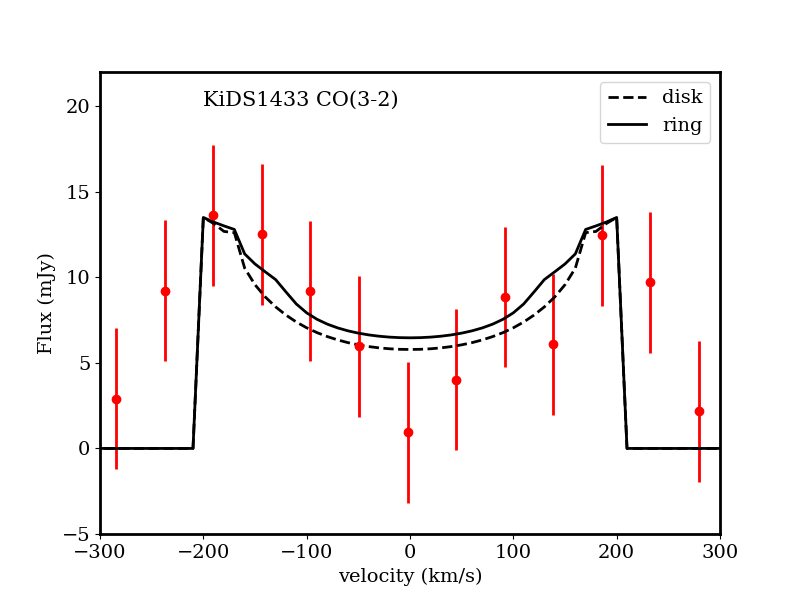}}
\caption{Modeling of the double-horn emission lines for the KiDS~1433. (a) $H_2$ gas density profile. (b) Rotational velocity curve. (c,d) Observed CO(1$\rightarrow$0) and CO(3$\rightarrow$2) spectra and modeling. {Red dots and error bars are the observed fluxes and their uncertainties, respectively. Dashed and solid black lines correspond to the disk and ring models, respectively.}  See the legend and text for further details.}\label{fig:double_horn_modeling}
\end{figure*}

\subsection{Double-horn modeling for KiDS~1433. Structural parameters.}\label{sec:double_horn_modeling}
In this section, we investigate the structural properties of KiDS~1433, which we have clearly detected in the first and third CO transitions. Interestingly, as discussed in Sect.~\ref{sec:IRAM30m_analysis}, this BCG shows evidence of a double-horn profile in both CO(1$\rightarrow$0) and CO(3$\rightarrow$2). As pointed out in C22a for the similar case of KiDS~1220, these double-horn spectra are indicative of a low concentration of the molecular gas toward the center of the galaxy, which causes the observed flux to dim at low velocities in Fig.~\ref{fig:BCG_spectra}. 

In the following, similarly to previous studies \citep{Wiklind1997,Salome2015}, we attempt to link the observed double-horn spectra to the structural parameters of the gas reservoir. To do this, we followed the procedure described in Sect.~5.5 of C22a and references therein, to which we refer for further details.

The observed velocity-integrated $S_{\rm CO(J\rightarrow J-1)}\Delta\varv$ spectrum 
can be related to the $H_2$ gas density profile $n(r)$ and the rotational velocity $V_{\rm rot}(r)$ using Eq.~4 of \citet{Wiklind1997},
\begin{equation}
\label{eq:model_double_horn}
 S_{\rm CO(J\rightarrow J-1)}\Delta\varv \propto \int\int \frac{n(r) r}{V_{\rm rot}(r)\sin i \sqrt{1-\bigg(\frac{\varv}{V_{\rm rot}(r)\sin i}\bigg)^2}}~dr~d\varv\,,
\end{equation}
where $r$ is the projected radius from galaxy center, and $i$ is the inclination with respect to the line of sight ($i = 0$~deg means face–on). We then parameterized the gas density profile with a \citet{Toomre1963} disk of order $n=2$, which corresponds to a flat Plummer distribution, parameterized by a characteristic disk scale $d$ and the central gas density $n_0=\frac{3 M_{\rm H_2}}{2\pi d^2}$, whose value is fixed by {matching} the surface integral of $n(r)$ to the total gas mass $M_{H_2}=5.9\times10^{10}~M_\odot$.

As an alternative to the flat disk, we considered an even less centrally concentrated distribution of $H_2$ as described by a ring. Following \citet{Wiklind1997}, we modeled a ring-like distribution as the difference of two gaseous disks with characteristic scales $d_1>d_2$. Similarly to the disk model, the characteristic density $n_0=\frac{3 M_{\rm H_2}}{2\pi (d_1^2-d_2^2)}$ was fixed by the surface integral.

We then expressed the total rotational velocity as the quadratic sum of the contributions from the gaseous component (disk or ring), the stellar bulge, and the dark matter (DM) halo.
Following \citet{Wiklind1997}, we modeled the stellar bulge contribution as follows:\begin{equation}
V_{\rm rot,bulge}(r)^2 =\frac{GM_\star r}{\left(r+\frac{r_{\rm eff}}{1.8153}\right)^2}\,,
\end{equation}
where $M_\star=1.86\times10^{11}~M_\odot$ (Table~\ref{tab:BCG_properties}) and $r_{\rm  eff}=7.47$~kpc are the stellar mass  and half-light radius of 
the KiDS~1433 BCG, respectively. We estimated the latter in \textsf{r} band using the {\sc SourceXtractor++} (v 0.16) package\footnote{\url{https://astrorama.github.io/SourceXtractorPlusPlus}}, which is the successor to the tool {\sc SeXtractor}  \citep{Bertin_Arnouts1996}.  We also derived  a  position-dependent model of the point spread function (PSF) with the {\sc PSFEx} tool\footnote{\url{www.astromatic.net}}, which we used as input in {\sc SourceXtractor++} to fit the galaxy with a S\'{e}rsic profile plus the model PSF. This analysis yielded $r_{\rm  eff}=(7.47\pm0.40)$~kpc, which agrees fairly well with that inferred from the mass versus size relation of galaxies at similar redshifts \citep{vanderWel2014}, and a S\'{e}rsic index $n_{\textrm{S\'ersic}}=3.05$. {The latter is lower than the characteristic $n_{\textrm{S\'ersic}}=4$ of elliptical galaxies, but also within the broad range of values ($n_{\textrm{S\'ersic}}\sim1-10$) that has been found for distant BCGs in previous studies \citep[e.g.,][]{Bai2014,Chu2021}.}

Following C22a, we then expressed the DM halo contribution as in Eq.~5 of \citet{Castignani2012}, which corresponds to the  universal rotation curve halo model by \citet{Salucci2007}.  In order to reproduce a flattening of the rotational velocity profile at a large distance $r\sim100$~kpc, as is the case for massive galaxies such as KiDS~1433 BCG, we assumed a DM mass of $M_{\rm DM}=4.7\times10^{12}~M_\odot$, which corresponds to a   stellar to DM halo mass ratio of $M_\star/M_{\rm DM}=4\%$ \citep[see, e.g.,][]{Girelli2020}, and implies a flattening of the rotation velocity to a value of $\sim300$~km~s$^{-1}$, which is typical of massive ellipticals such as KiDS~1433.

In order to determine the structural parameters of the $H_2$ reservoir, we then varied the disk scales $d$, $d_1$, and $d_2$ from 50~pc up to 20~kpc. By integrating Eq.~\ref{eq:model_double_horn} over all radii and over the velocity support of the CO(1$\rightarrow$0) and CO(3$\rightarrow$2) lines separately, we related the observed spectrum to the spectrum inferred from our modeling. 
The overall normalization was fixed by the height of the line. Increasing the inclination angle $i$ has the net effect of increasing the velocity separation between the two peaks in the spectrum \citep{Wiklind1997}. As $M_{H_2}$ is fixed, decreasing the disk scales $d$, $d_1$ has the net effect of increasing the characteristic density $n_0$, and thus the concentration, which ultimately reduces the observed depth in the spectrum at low velocities with respect to the peaks. Given these relations, we find that a ring and a disk with an inclination of $i=45$~deg {both well reproduce}  the observed spectrum. The inferred sizes are similar for the disk and ring models, $d=7$~kpc and $d_1=5$~kpc, respectively, while $d_1-d_2=1$~kpc. With these parameters, we obtain  peak $H_2$-gas-density values of $\sim$570 and 430~$M_\odot$/pc$^2$ in the case of a disk and a ring, respectively, which are typical values found for LIRGs, such as our BCGs,  along the Kennicutt-Schmidt relation.

Figure~\ref{fig:double_horn_modeling} displays the gas surface density profiles (a) and  the rotational velocity profile (b) in the case of a $d=7$~kpc disk, and similar curves are obtained in the case of a ring with the above parameters. In the bottom panels~(c) and (d) we instead show the observed CO(1$\rightarrow$0) and CO(3$\rightarrow$2) spectra, respectively, over which we plot our model curves.  The adopted set of parameters reproduces the CO(1$\rightarrow$0) and CO(3$\rightarrow$2) spectra well, which further supports the overall modeling. Interestingly, the star-forming KiDS~1220 BCG at $z=0.39$ shows a similar double-horn profile. Recently,
C22a modeled the spectrum with an  extended ($d=9$~kpc) and low-concentration molecular gas reservoir, similarly to {the} KiDS~1433 BCG. The two KiDS~1220 and KiDS~1433 BCGs also have similar stellar masses, $M_\star\simeq(2-3)\times10^{11}~M_\odot$, gas-to-stellar mass ratios $M_{H_2}/M_\star\simeq(0.3-0.5)$, and effective radii $r_{\rm eff}\simeq7$~kpc.

{Our results suggest that extended molecular gas reservoirs may be common in intermediate-$z$ and star-forming BCGs, as indeed two (i.e., 50\%) out of the four $z\sim0.4$ KiDS BCGs detected in CO by C22a and in this study exhibit double-horn CO spectra, well modeled by extended gas reservoirs. Similarly, by studying a sample of local ($z<0.017$) brightest group galaxies, \citet{Olivares2022} found that 39\% of their sources are dominated by extended disk and ring-like structures. 
These results may suggest that the star formation and gas properties of star-forming BCGs of moderately rich clusters and groups do not significantly evolve over 4~Gyr of cosmic time between $z\sim0$ and $z\sim0.4$. {This agrees with recent results by \citet{Orellana-Gonzalez2022}, who indeed showed that the average SFR of star-forming BCGs (i.e., those with SFR$>10~M_\odot$/yr) does not strongly evolve with redshift, within the interval $z\sim0-0.4$.} Regardless of the specific cosmic time, gas-rich BCGs are caught in a phase of significant star formation above the MS levels.
As suggested by \citet{Olivares2022}, during this phase, the angular momentum may prevent the cold gas from being rapidly accreted, which would explain the low concentration and the large extent of the molecular gas reservoirs.}

\section{Summary and conclusions}\label{sec:conclusions}

We have studied the molecular gas and star formation properties of three star-forming BCGs at intermediate redshifts $z\sim0.4$. We selected the sources as they exhibit significant infrared emission (Sect.~\ref{sec:BCGselection}) from a parent sample of 684 spectroscopically confirmed $z>0.3$ BCGs in the  equatorial KiDS-N field \citep{Radovich2020}. Together with three similarly star-forming $z\sim0.4$ KiDS BCGs drawn from the same parent sample as we studied in our recent C22a work, the targeted BCGs are among the most star-forming BCGs in KiDS.

The BCGs considered in this work have indeed line-based ([O~II]) and SED-based ${\rm SFR}\simeq20~M_\odot$/yr, on average, as well as dust masses and luminosities  typical of LIRGs (Sects.~\ref{sec:SEDs} and \ref{sec:line_diagnostics}). Similarly, on the basis of the position of the sources in the infrared WISE color-color plane \citep{Jarrett2017}, two BCGs are classified as starbursts and one as an intermediate disk, which confirms their significant ongoing star formation activity (Sect.~\ref{sec:IR_diagnostics}).

We investigated the stellar, star formation, and gas properties of the three BCGs with the {aim of better understanding} the star formation fueling and stellar mass assembly of distant BCGs. We performed an analysis of the color-magnitude plots for the cluster members (Sect.~\ref{sec:color_magnitude_plots}). Our red-sequence modeling favors a recent burst of star formation for the BCGs. For KiDS~1415, in particular, our modeling implies a remarkably recent (300~Myr) burst of star formation. This result, together with the high star formation activity and high infrared luminosity of the BCG, suggests that the source may be interacting with its nearby companion(s).


In addition, we searched for molecular gas feeding the star formation in the three KiDS BCGs of this work. To do this, we observed them with the IRAM 30m telescope at the first three CO transitions (Sect.~\ref{sec:observations_and_data_reduction}). We clearly detected double-horn CO(1$\rightarrow$0) and CO(3$\rightarrow$2)  emission lines for one of our targets, that is, {the} KiDS~1433 BCG ($z=0.3546$). Our observations and fits of the CO lines imply a large molecular gas reservoir of $M_{H_2}=(5.9\pm1.2)\times10^{10}~M_\odot$ and a high gas-to-stellar mass ratio $M_{H_2}/M_\star=(0.32^{+0.12}_{-0.10})$. We thus increase the still limited sample of distant BCGs with detections in multiple CO transitions. The double-horn emission for KiDS~1433 implies a low gas concentration, while a modeling of the spectra yields an extended molecular gas reservoir, with a characteristic radius of $\sim$(5-7)~kpc (Sect~\ref{sec:double_horn_modeling}), which is reminiscent of the mature extended-disk phase observed in some local BCGs \citep{Olivares2022}, as well as in the star-forming KiDS~1220 BCG  ($z=0.3886$, C22a). These results suggest that extended molecular gas reservoirs may be common in intermediate-$z$ and star-forming BCGs.

For the other two KiDS BCGs that we observed with the IRAM-30m telescope in this work, we were able to set robust upper limits of $M_{H_2}/M_\star<0.07$ and $<0.23$, which are among the lowest for distant BCGs. We then combined our observations with available stellar, star formation, and dust properties of the targeted BCGs, and compared them with a sample of $\sim100$ distant cluster galaxies (Sect.~\ref{sec:comparison_sample}), including additional intermediate-$z$ BCGs, with observations in CO from the literature. In particular, we confirm at higher significance than in our recent C22a work the correlation between the excitation ratio, the sSFR, and the star formation efficiency, that is, the inverse of the depletion time, separately (Sect.~\ref{sec:excitation_ratios}).

In summary, we suggest that the intermediate-$z$ KiDS BCGs that are detected in CO, which have high $\tau_{\rm dep}$, $M_{H_2}/M_\star$, and   $M_{H_2}/M_{\rm dust}$, are observed in a peculiar and rare phase of the BCG evolution, in which a substantial amount of the $H_2$ gas has recently been accreted, but still not efficiently converted into stars. As has been suggested in C22a, the accretion may have occurred   via interaction with the cluster companions.  The other star-forming BCGs (KiDS~0837 and 1415), with upper limits in CO alone, may be at a different and possibly later stage of the BCG evolution, in which the $H_2$ gas reservoir has been consumed or is undergoing rapid consumption on a timescale $\tau_{\rm dep}\lesssim(1-2)$~Gyr, while star formation is still ongoing. These results suggest that molecular gas depletion precedes star formation quenching in intermediate-$z$ and star-forming BCGs (Sect.~\ref{sec:thoughts_gas_BCGs}). This result agrees with what \citet{Castignani2022c} found locally for early-type galaxies in the field of the Virgo cluster.

\begin{acknowledgements}
{ We thank the anonymous referee for helpful comments which contributed to improving the paper.}
This work is based on observations carried out under project numbers 176-21 with the IRAM 30m telescope. The research leading to these results has received funding from the European Union’s Horizon 2020 research and innovation program under grant agreement No 101004719 [ORP]. GC thanks IRAM staff at Granada for their help with these observations. IRAM is supported by INSU/CNRS (France), MPG (Germany) and IGN (Spain). 
GC, LM, CG, and FM acknowledge the support from the grant ASI n.2018-23-HH.0.
LM and CG acknowledge the support from the grant PRIN-MIUR 2017 WSCC32. CG acknowledges funding from the Italian National Institute of Astrophysics under the grant "Bando PrIN 2019", PI: Viola Allevato and
from the HPC-Europa3 Transnational Access programme HPC17VDILO.
MS acknowledges financial contribution from contracts ASI-INAF n.2017-14-H.0 and INAF mainstream project 1.05.01.86.10.
PS acknowledges support by the ANR grant LYRICS (ANR-16-CE31-0011).
DT acknowledges the support from the Chinese Academy of Sciences (CAS) President's International Fellowship Initiative (PIFI) with Grant N. 2020PM0042.
The research has made use of the NASA/IPAC Extragalactic Database (NED), which is operated by the Jet Propulsion Laboratory, California Institute of Technology,
under contract with the National Aeronautics and Space Administration.
The work uses GAMA data-products. GAMA is a joint European-Australasian project based around a spectroscopic campaign using the Anglo-Australian Telescope. The GAMA input catalogue is based on data taken from the Sloan Digital Sky Survey and the UKIRT Infrared Deep Sky Survey. Complementary imaging of the GAMA regions is being obtained by a number of independent survey programs including GALEX MIS, VST KiDS, VISTA VIKING, WISE, Herschel-ATLAS, GMRT and ASKAP providing UV to radio coverage. GAMA is funded by the STFC (UK), the ARC (Australia), the AAO, and the participating institutions. The GAMA website is \href{http://www.gama-survey.org/}{http://www.gama-survey.org/}. The work is based on data products from observations made with ESO Telescopes at the La Silla Paranal Observatory under program IDs 177.A-3016, 177.A-3017 and 177.A-3018, and on data products produced by Target/OmegaCEN, INAF-OACN, INAF-OAPD and the KiDS production team, on behalf of the KiDS consortium. OmegaCEN and the KiDS production team acknowledge support by NOVA and NWO-M grants. Members of INAF-OAPD and INAF-OACN also acknowledge the support from the Department of Physics \& Astronomy of the University of Padova, and of the Department of Physics of Univ. Federico II (Naples). 
\end{acknowledgements}

\end{document}